\documentclass{JHEP3}
\usepackage{amsmath}
\usepackage{axodraw}
\usepackage{epsfig}
\usepackage{amssymb}
\usepackage{mathrsfs}
\usepackage{mathcomp}
\usepackage{cite}
\usepackage[latin1]{inputenc}

\def\rap{\ensuremath{Y}}

\keywords{Dipole Model, Small-$x$ evolution, Saturation}
\preprint{LU-TP 07-09\\
}

\title{On the Dipole Swing and the Search for Frame 
  Independence in the Dipole Model}

\author{Emil Avsar\\
  Dept.~of Theoretical Physics,
  Sölvegatan 14A, S-223 62  Lund, Sweden\\
  E-mail: \email{emil@thep.lu.se}}

\abstract{Small-$x$ evolution in QCD is conveniently described by 
  Mueller's dipole model which, however, does not include saturation effects 
  in a way consistent with boost invariance. 
  In this paper we first show that
  the recently studied zero and one dimensional toy models exhibiting saturation
  and explicit boost invariance can be interpreted in terms positive definite 
  $k\to k+1$ dipole vertices. Such $k\to k+1$ vertices can in the full model 
  be generated by combining the usual dipole splitting with $k-1$ simultaneous dipole 
  swings. We show that, for a system consisting of $N$ dipoles, one needs to 
  combine the dipole splitting with at most $N-1$ simultaneous swings in 
  order to generate 
  all colour correlations induced by the multiple dipole interactions.  
}

\begin{document}

\sloppy

\section{Introduction}

The small-$x$ region in QCD can be described by the well known,
linear, BFKL equation which predicts a power like growth in $x$ for the gluon density. 
Such a fast growth is problematic since it breaks the unitarity bound 
at high energies. As the gluon density becomes large, non-linear effects cannot be ignored 
and it was early suggested that effects from parton saturation should tame the 
growth of the gluon density, in accordance with unitarity \cite{Gribov:1984tu}.

Starting the evolution from a colour singlet quark-antiquark pair, a colour dipole, 
Mueller \cite{Mueller:1993rr, Mueller:1994gb}
formulated a dipole model in transverse coordinate space which reproduces 
the BFKL equation to leading order. The transverse coordinate formulation 
also allows one to go beyond the BFKL equation since it is here easier 
to take into account multiple interactions. This is so because the transverse 
coordinates of the partons are frozen during the evolution, and it is therefore 
rather easy to sum the multiple scattering series in an eikonal approximation. 
This was exploited by Mueller, who was thus able to obtain a unitarised formula 
for the scattering amplitude.  

Within the dipole formalism, Balitsky \cite{Balitsky:1995ub} 
derived an infinite
hierarchy of equations for the dipole scattering amplitudes.  Kovchegov \cite{Kovchegov:1999yj} 
derived a closed equation for the amplitude using a mean field approximation, and this 
equation is referred to as the 
Balitsky-Kovchegov (BK) equation. 
The same hierarchy of equations also follows from the JIMWLK equation
\cite{Jalilian-Marian:1997jx,Jalilian-Marian:1997gr,Iancu:2001ad,Weigert:2000gi}
which is the master equation of the Color Glass Condensate (CGC) formalism 
\cite{Iancu:2002xk} 
and describes the non-linear evolution of dense hadronic matter in the presence of saturation 
effects. 

%Summing multiple scatterings to all orders, one can in Mueller's model 
%obtain a unitarised expression for the scattering amplitude. 

In Mueller's model the multiple dipole interactions correspond to 
multiple pomeron exchange, and in the Lorentz frame where the collision is 
studied these multiple interactions lead to the formation of pomeron loops. 
However, these loops cannot be formed during the 
evolution of the dipole cascade since this evolution is linear. Thus only 
those loops which are cut in the specific Lorentz frame used for the calculation
are accounted for, while none of the remaining loops is included. This implies 
that the model is not frame independent. To minimize the error, the optimal frame to use is the 
one where the colliding dipole cascades are of the same density, since multiple 
scatterings then become important  at rapidities where one may still neglect saturation 
effects in the evolving dipole cascades.  

In order to obtain a frame independent formalism it is necessary to 
include saturation effects also in the evolution of each dipole cascade. 
There have been various attempts to include such saturation effects 
in a consistent manner, but no explicitely frame independent formalism 
has yet been presented.
%There 
%have been attempts to interpret such effects as being due to dipole mergings, 
%which require a $2\rightarrow 1$ transition vertex. 
%It was shown in \cite{Iancu:2005dx} that such a vertex 
%is not positive definite, and therefore one cannot give a meaningful probabilistic 
%interpretation for the process. Another approach was taken in \cite{Kozlov:2006cg} where 
%the authors calculated directly the $4 \to 2$ gluon merging vertex within the 
%dipole language and they were led to the conclusion that one should include 
%a $2\to 4$ dipole vertex to take this process into account. However, no explicitely 
%frame independent formalism has yet been presented, in which the evolution can be 
%described as a simple semiclassical cascade. 

%\cite{Salam:1996nb,Salam:1995uy, Avsar:2005iz, Avsar:2006jy, Avsar:2004ms, Avsar:2007ht}. 

To gain insight and possible hints towards a solution, a simple 1+0 dimensional (rapidity 
constitutes the only dimension) toy model in which transverse 
coordinates are neglected was constructed in \cite{Mueller:1994gb, Mueller:1996te}. This model has  
been further 
studied in \cite{Blaizot:2006wp, Kovner:2005aq}, and the resulting frame-independent evolution 
can be interpreted as a coherent emission of new ``dipoles'' from the multi-''dipole'' state. 

In \cite{Avsar:2005iz} we developed a dipole cascade model for DIS and 
$pp$ collisions based on 
Mueller's model but also including effects of energy-momentum conservation. 
This model was further extended in \cite{Avsar:2006jy}
to include saturation effects in the cascade evolution through the so called dipole swing mechanism 
\cite{Kozlov:2006cg, Avsar:2006jy}, which gives an additional $2\to 2$ transition 
during the evolution. Monte Carlo (MC) simulations show that the evolution is almost frame 
independent, and the model results are in good agreement with inclusive and 
diffractive data from HERA and the Tevatron \cite{Avsar:2007xg}. 

In this paper we will first show that the explicitely frame independent evolution in the 
toy model mentioned above, and also in its 1+1 dimensional generalization in 
\cite{Iancu:2006jw}, can be given a probabilistic interpretation in terms of positive 
definite $k \to k+1$ dipole vertices. 
Such $k \to k+1$ transitions can in the full model be generated by combining the 
dipole splitting with the dipole swing. 

In case each individual dipole is restricted to single scattering only, 
we show that one needs to combine the dipole splitting with at most 
one swing at a time in order to generate the necessary colour 
correlations. As remarked in \cite{Blaizot:2006wp}, the toy 
model evolutions mentioned above describe the multiple scatterings
of individual dipoles, and we will here show that one can in 
the real model generate the correlations induced by the dipole scatterings
by combining a splitting with several simultaneous swings. 
For a system consisting of $N$ spatially uncorrelated dipoles, it is easy
to see that 
one needs to combine the dipole splitting with at most $N-1$ 
simultaneous swings. In a process where one splitting is combined 
with $k-1$ swings, $k$ dipoles are replaced by $k+1$ dipoles, thus 
giving a $k\to k+1$ transition. 

However, starting the evolution from a single $q\bar{q}$ pair,
one obtains dipoles which are connected in chains, and in this 
case not all swings are allowed. It is here important to keep 
track of the correct topology of the dipole state. While this is 
never a problem in the original formulation which only contains 
the $1\to 2$ splitting, it is here very important  
to avoid the formation of unphysical states. 
Although it has been checked for a large number of cases, 
a formal proof that it is always enough with $N-1$ swings 
is not avaliable, and the result is a conjecture. 

%The interpretation of the saturation mechanism in the toy models within 
%the dipole picture is, however, somewhat unclear, and it is different from the 
%saturation mechanism presented in \cite{Avsar:2006jy}. These points are 
%further discussed in the forthcoming sections.  

The paper is organized as follows. In the next section we shortly review
Mueller's dipole model and the question of frame independence 
in the evolution. In section \ref{sec:toymodels} we will review the 
toy models in zero and one transverse dimensions formulated in \cite{Mueller:1996te, Iancu:2006jw}, 
and show how these can be interpreted in terms of the $k \to k+1$ transitions 
mentioned above.  
Then in section \ref{sec:fulldipmod} we consider the evolution in the 
full model and argue that also in this case the correct evolution can 
be formulated in terms of $k\to k+1$ vertices. In section \ref{sec:evtopology} we go on 
to study the colour topology of the evolution, and we 
show how one can generate the 
needed colour correlations using the dipole swing. Finally, 
in section \ref{sec:conclusions}, we present our conclusions. 

\section{Approaches Towards a Frame Independent Formalism}
\label{sec:dipmod}

In Mueller's model\cite{Mueller:1993rr,Mueller:1994gb} for onium-onium scattering a 
colour dipole formed by a colour 
charge at transverse coordinate $x$ and an anti-charge at $y$ can 
split into two dipoles by emitting a soft gluon at position $z$ with the 
following probability 
\begin{eqnarray}
\frac{d\mathcal{P}}{d\rap}&=&\frac{\bar{\alpha}}{2\pi}d^2 z
\frac{(x-y)^2}{(x-z)^2 (z-y)^2}
\equiv d^2z \mathcal{M}(x,y,z), 
\label{eq:dipkernel} \\
\bar{\alpha} &\equiv& \frac{\alpha_sN_c}{\pi}\, \, \, \mathrm{and}
\, \, \, Y \equiv \mathrm{ln}\frac{1}{\mathrm{x}_{Bj}}. \label{eq:alphabar}
\end{eqnarray}
We refer to $\mathcal{M}$ as the dipole kernel and to $Y$ as the rapidity, which here 
acts as the time variable in which the evolution proceeds. A dipole cascade is then formed
when each dipole splits repeatedly. When two such cascades collide, a right-moving dipole
$(x_i,y_i)$ and a left-moving dipole $(x_j,y_j)$ interact 
with a probability 
\begin{equation}
f_{ij} = f(x_i,y_i|x_j,y_j) = \frac{\alpha_s^2}{8}\biggl[\log\biggl(
\frac{(x_i-y_j)^2
(y_i- x_j)^2}{(x_i-x_j)^2(y_i-y_j)^2}\biggr)\biggr]^2.
\label{eq:dipamp}
\end{equation}
All dipole interactions are assumed to be independent, and the $S$-matrix element is 
given by $S=$ exp$(-\sum_{ij}f_{ij})$. 
%Therefore the probability for no interaction is given by
%\begin{equation}
%P(\mathrm{no} \, \, \mathrm{interaction}) = S^2 = \mathrm{exp}(-2\sum_{ij}f_{ij}).
%\label{eq:Pnotint}
%\end{equation} 
%The dipoles which do not interact cannot come on shell and consequently they cannot appear 
%in the final state and must be treated as virtual. Thus in this formalism the 
%saturation need not to decrease the total number of dipoles but rather only 
%the number of ``real'' or interacting dipoles. 

In this formalism, saturation effects occur only due to multiple scatterings while 
the evolution of the dipole cascade satisfies the usual BFKL equation. This 
implies that the formalism is not frame independent, and in order to obtain a frame 
independent formalism such saturation effects must be properly included in the cascade 
evolution. 

Different approaches have been proposed to obtain this. 
It was noted that the CGC formalism 
is not complete in the sense that it does not contain any gluon splittings, or gluon number 
fluctuations. This problem comes from the fact that the Balitsky-JIMWLK (B-JIMWLK) 
equations\footnote{Throughout this 
paper, we will only consider the large-$N_c$ version of these equations.}, which can 
schematically be written as 
\begin{equation}
\partial_Y \langle T^k \rangle = \mathcal{M} \otimes \{ \langle T^k \rangle -  \langle T^{k+1} \rangle \},
\label{eq:bjimwlk}
\end{equation}
only couples the $k-$dipole scattering amplitude $T^k$ to the $(k+n)$-dipole amplitudes with 
$n=0, 1, \dots$. From the view of target\footnote{In the CGC approach, it is usually 
assumed that the target is a dense hadron while the projectile is an elementary colour dipole.} 
evolution, this means 
that one includes all gluon merging diagrams, while the gluon splitting diagrams are absent. 
Gluon splittings are equivalent to dipole splittings, and the dipole model has been used to 
add fluctuations into the formalism. The modified B-JIMWLK equations then read
\begin{equation}
\partial_Y \langle T^k \rangle = \mathcal{M} \otimes \{ \langle T^k \rangle -  \langle T^{k+1} \rangle \}
+ \mathscr{K} \otimes \langle T^{k-1} \rangle,
\label{eq:modbjimwlk}
\end{equation}
where $\mathscr{K}$ is a kernel representing the fluctuations in the target. 

Viewed in the opposite direction, the fluctuation effects in the target correspond 
to saturation effects in the projectile evolution. Given the form of the modified B-JIMWLK 
equations above, it may seem natural to include a $2\to 1$ vertex in the dipole evolution.
Such an interpretation has, however, the drawback 
that it cannot be interpreted as a classical evolution process since the $2\to 1$ vertex is 
not positive definite, as was shown by Iancu et al. \cite{Iancu:2005dx}. 
In \cite{Kozlov:2006cg} 
Kozlov et al. calculated directly the $4 \to 2$ gluon merging vertex within the 
dipole language, and were led to the conclusion that one should include 
a $2\to 4$ dipole vertex which is composed of a splitting and a swing. 

%Indeed, notice that the fluctuation term only 
%appears as the $\langle T^{k-1} \rangle$ contribution, 
%while the $\langle T^k \rangle$ contribution is unaffected by this vertex.
%In \cite{Kozlov:2006cg} 
%the authors calculated directly the $4 \to 2$ gluon merging vertex within the 
%dipole language and they were led to the conclusion that one should include 
%a $2\to 4$ dipole vertex to take this process into account.However, no explicitely 
%frame independent formalism has yet been presented.  

%This seems 
%a little odd since one would expect such a vertex also to give a 
%contribution to the $\langle T^k \rangle$
%term as well (this will be commented more in section \ref{sec:eveq}). 

In \cite{Avsar:2006jy} we included the $2\to 2$ dipole 
``swing'',
in addition to the $1\to 2$ dipole splitting, in the evolution. 
If we have two dipoles $(x_i,y_i)$
and $(x_j,y_j)$, the swing will replace them by $(x_i,y_j)$ and 
$(x_j,y_i)$.
The dipole swing can be interpreted in two ways. First, as a way 
to approximate colour quadrupoles as two independent dipoles formed by the closest 
charge--anti-charge pairs, in which case the swing is naturally suppressed by $N_c^2$. Secondly, 
we may view it as the result of a gluon exchange between the dipoles, 
which results in a change in the colour flow. In this case the swing would be proportional 
to $\alpha_s^2$, which again, compared to $\bar{\alpha}$, is formally suppressed by $N_c^2$.   

The dipole swing is related to the pomeron interactions studied by Bartels and Ryskin 
\cite{Bartels:1993it, Bartels:1993ke}. 
Here a pomeron is interpreted as two gluons in a colour singlet state. In a four gluon 
system with two singlet pairs, gluon exchange can give a transformation ( a ``switch'' 
or ``swing'') $(12)+(34)\to (13)+(24)$, where a parenthesis denotes two gluons in a 
singlet state. 

We note that the swing is here not a 
vertex in the same sense as the splitting process since, unlike the splitting, the swing 
is not proportional to $dY$ but rather happens instantaneously. In the MC implementation we assign a 
``colour'' to each dipole 
and two dipoles are allowed to swing if their colour indices match\footnote{The number 
of effective colours is $N_c^2$ which is the number of possible colour configurations 
for a given colour--anti-colour pair.}. 
The swing is then determined by the weight 
\begin{eqnarray}
P(\mathrm{swing}) \propto  
\frac{(x_1-y_1)^2(x_2-y_2)^2}{(x_1-y_2)^2(x_2-y_1)^2}.
\label{eq:probswing}
\end{eqnarray}
Here the two initial dipoles are determined by the coordinates $(x_1,y_1)$ 
and $(x_2,y_2)$, and the new by $(x_1, y_2)$ and $(x_2, y_1)$. 
This form favours the formation of small dipoles. It also preserves one of the results in 
Mueller's original formulation, namely that the total weight for a dipole chain is given by 
the product $\prod_i \frac{1}{r_i^2}$, where 
$r_i$ is the size of dipole $i$ and the product runs over all ``remaining'' dipoles 
in the cascade. 

In our formalism
the total number of dipoles does not decrease. For each event, many of the dipoles 
will not interact and these have to be considered as being virtual.  
In this case saturation effects do not have to decrease the total number of dipoles but rather
only the number of interacting, or ``real'', dipoles. The dipole swing 
has this property since it  
is more likely that two dipoles are replaced by two smaller dipoles, as can easily be 
seen from \eqref{eq:probswing}, and smaller dipoles  
have smaller interaction probabilities. Thus the number of interacting dipoles will actually
decrease, and in pomeron language this means that the swing generates pomeron 
mergings. 

An essential feature of our formalism is that the dipoles in the cascade form 
connected chains\footnote{In Mueller's original formulation this is not relevant since 
  there the dipoles evolve truly independently. However, in our implementation of energy-momentum 
  conservation, neighboring dipoles affect each other and it is then relevant that 
  the cascade is formulated as a dipole chain.}, 
rather than a collection of uncorrelated dipoles, as in a reaction-diffusion 
type of formalism. A dipole chain cannot end in a gluon, and it is not possible to remove 
a dipole without reconnecting its neighbors. A generic $2\to 1$ vertex is therefore 
not possible in this formalism\footnote{The only allowed merging process 
is in case two neighboring dipoles merge.}, and the dipole swing gives the simplest process from 
which one can form closed chains during the evolution.

\section{The Toy Models}
\label{sec:toymodels}

\subsection{The 1+0 dimensional toy model}
\label{sec:zerotoymod}

In this section we will review the toy model which 
was studied in detail in \cite{Blaizot:2006wp} (see also \cite{Kovner:2005aq}). 
This model was first presented by Mueller \cite{Mueller:1994gb} and it is interesting since, 
besides having some structural aspects in common with the dipole model,  
it offers analytical solutions which have been very hard to obtain 
for the full model. 

The model is defined such that at any rapidity $Y$ the system is specified only 
by the number of dipoles.  The probability to find the system
in the $n$-dipole state at time $Y$ is denoted by $P_n(Y)$. Here transverse 
coordinates are completely neglected and $Y$ defines the 
only coordinate in the model.  

If we let $\pmb{\mathcal{H}}$ denote the Hamiltonian of the system we 
have  
\begin{equation}
\mathcal{H}_{nm}=\langle m|\pmb{\mathcal{H}} |n\rangle =
\mathcal{R}(n)(\delta_{m,n+1}-\delta_{m,n}).
\label{eq:Fzerodim}
\end{equation}
$\mathcal{R}(n)$ is so far 
an unspecified function, which determines the splitting rate of the $n$ 
dipole state.
The probability $P_n(Y)$ evolves according to 
\begin{eqnarray}
\partial_Y P_n = \mathcal{H}_{mn}P_m.
\label{eq:pnevolution}
\end{eqnarray}
%For any $Y$, the probabilities must satisfy the condition (we set $P_0=0$) 
%\begin{eqnarray}
%\sum_{n=1}^\infty P_n(Y) = 1
%\end{eqnarray}
%It is immediately seen
%that the evolution \eqref{eq:pnevolution}, with $\pmb{\mathcal{F}}_{nm}$ given by
%\eqref{eq:Fzerodim}, preserves this condition. 

With $S_{nm}$ we denote the $S$-matrix for the scattering of two dipole 
states of $n$ and $m$ dipoles respectively. If we assume that each dipole scatters 
independently with a probability $\tau$, we have $S_{nm}=(1-\tau)^{nm}$. We here assume 
that $\tau << 1$ and if this is not the case, one should replace $1-\tau$ by 
exp$(-\tau)$. The physical $S-$matrix, $\mathcal{S}$, is obtained by taking an 
average over all possible events. Using matrix notation we have
\begin{equation}
\mathcal{S}(Y_1+Y_2)=\pmb{p}^T(Y_1)\pmb{S}\pmb{q}(Y_2).
\label{eq:toySmatrix}
\end{equation}   
Here $\pmb{p}^T(Y_1)=(P_1(Y_1), P_2(Y_1), \dots )$ is the row vector of the configuration
probabilities for the right moving onium (evolved up to $Y_1$) while $\pmb{q}$ denotes 
the column vector of configuration probabilities for the left moving onium 
(evolved up to $Y_2$). 

In \eqref{eq:toySmatrix} we have anticipated that $\mathcal{S}$ depends only  
on the total rapidity interval $Y_1+Y_2$, which defines boost invariance. 
This implies that we have $(\partial_{Y_1}-\partial_{Y_2})\mathcal{S}=0$, and 
requiring this in \eqref{eq:toySmatrix} one obtains 
\begin{equation}
\pmb{p}^T(Y_1)\pmb{\mathcal{H}}\pmb{S}\pmb{q}(Y_2) - 
\pmb{p}^T(Y_1)\pmb{S}\pmb{\mathcal{H}}^T\pmb{q}(Y_2)
= 0.
\end{equation}
A sufficient condition for a solution is to require that 
\begin{equation}
\pmb{\mathcal{H}}\pmb{S}=\pmb{S}\pmb{\mathcal{H}}^T
\label{eq:boostinv}
\end{equation}
which means that $\pmb{\mathcal{H}}\pmb{S}$ is symmetric
(since $\pmb{S}$ is symmetric). It is now easily seen that 
condition \eqref{eq:boostinv} requires $\mathcal{R}(n)$ in \eqref{eq:Fzerodim} to be given by 
\begin{equation}
\mathcal{R}(n)=c\,(1-(1-\tau)^n)
\label{eq:toyf}
\end{equation}
where $c$ is a constant, $c=\mathcal{R}(1)/\tau$. By rescaling $Y$, 
we might as well assume $c$ to equal 1. 

%The total splitting rate in \eqref{eq:toyf} describes the coherent emission of 
%the new particle. We see that $\mathcal{R} \rightarrow c$ as $n \rightarrow \infty$, so that 
%the splitting rate saturates as the system gets very dense. This 
%is in contrast to the original dipole model where each dipole 
%can emit a gluon independently from the rest of the system, which 
%implies that $\mathcal{R} \rightarrow \infty$ as $n\rightarrow \infty$. Thus 
%saturation is here a coherent effect which reduces the total splitting rate 
%for very dense systems. In \cite{Blaizot:2006wp} the process is interpreted 
%as multiple scatterings between  the newly emitted particle and the $n$ particle 
%state from which it was emitted. This is indeed similar to what happens in the CGC formalism.

\subsection{Stochastic evolution with $k\to k+1$ vertices}
\label{sec:stochmod}

We note that \eqref{eq:toyf} can be rewritten as
\begin{equation}
\mathcal{R}(n)=c\sum_{k=1}^n \binom{n}{k}(-1)^{k-1}\tau^k.
\label{eq:toyfk}
\end{equation}
This suggests that we can interpret the evolution in terms of $k\to k+1$ 
transitions with weights $(-1)^k \tau^k$. 
However, the alternating signs implies that one cannot interpret 
these vertices in a probabilistic formulation. 

We will now show that one can nevertheless interpret the 
evolution in terms of positive definite $k\to k+1$ vertices. 
Thus the evolution can still be formulated as a stochastic 
process, but we will also see that the probabilistic interpretation 
of the evolution implies that it cannot be reduced 
into a formalism which describes a system of incoherent particles.   

Assume we have a system of $n$ particles $X$, which we also refer to 
as dipoles, satisfying the following rules. Each isolated $X$ can 
emit a new $X$ with a probability per unit time (rapidity) given by $\tau$, \emph{i.e.} 
we have a reaction $X\to X+X$ which occurs with probability $\tau > 0$. 
In addition to this, $k$ isolated dipoles can undergo a 
transition $kX\to (k+1)X$ with probability $\tau^k$. 
 
The evolution of the probabilities $P_n(Y)$ for these $X$ then satisfies
\begin{equation}
\partial_YP_n(Y) = -\sum_{k=1}^{n} \mathcal{R}^{(n)}_{k\rightarrow k+1}
P_n(Y) + \sum_{k=1}^{n-1} \mathcal{R}^{(n-1)}_{k\rightarrow k+1} P_{n-1}(Y)
\end{equation}
where $\mathcal{R}^{(n)}_{k\rightarrow k+1}$ is the splitting rate for the 
process where $k$ dipoles are replaced by $k+1$ dipoles in a state containing $n$ dipoles.

Now, the splitting rate $\mathcal{R}_{k\to k+1}^{(n)}$ is not simply given by $\binom{n}{k}\tau^k$ 
as one could expect naively, but it is instead given by 
\begin{equation}
\mathcal{R}^{(n)}_{k\rightarrow k+1} = \binom{n}{k} \tau^{k}(1-\tau)^{n-k},
\label{eq:probtoyfk}
\end{equation}
since for each $k\to k+1$ we must also multiply with the probability that 
no more than $k$ dipoles were involved in the emission of the new dipole. Obviously,
$k$ must run from 1 to $n$, and summing all contributions we obtain 
the total splitting rate as 
\begin{eqnarray}
\sum_{k=1}^n \mathcal{R}_{k\to k+1}^{(n)} = \sum_{k=1}^n \binom{n}{k}\tau^{k}
(1-\tau)^{n-k}=(1-(1-\tau)^n) = \mathcal{R}(n)
\end{eqnarray}
where $\mathcal{R}(n)$ was defined in \eqref{eq:toyf}. The positive definite 
transition rates in \eqref{eq:probtoyfk} thus give a boost invariant 
evolution as before. Note also that in this case the $k\to k+1$ transitions are not
universal since they depend on $n$, unlike the rates in \eqref{eq:toyfk}. 
The $k\to k+1$ splitting vertex therefore not only depends on the 
state of the $k$ emitters, but it does also depend on the rest of the dipoles 
in the cascade. We will in the forthcoming sections see that 
the dipole swings give a very similar evolution in the full model. 

We also note that the $k \to k+1$ transitions can be made manifest by 
writing down the Hamiltonian 
\begin{eqnarray}
\pmb{\mathcal{H}} = \sum_{k=1}^\infty \frac{\tau^k}{k!} ( \pmb{N}^{-1/2} \pmb{a}^\dagger - \pmb{1}) 
(\pmb{a}^\dagger)^k\pmb{a}^k \prod_{l = k+1}^\infty (1-\tau) 
\sum_{m=l}^\infty |m\rangle \langle m|,
\end{eqnarray}
where $\pmb{a}^\dagger$ and $\pmb{a}$ are dipole creation and annihilation operators 
respectively, and $\pmb{N} =\pmb{a}^\dagger\pmb{a}$.

\subsection{Evolution equations}
\label{sec:eveq}

In this section we will first show that the evolution 
equations  for the scattering amplitudes
derived in \cite{Blaizot:2006wp} are described \emph{exactly} by the 
$k \to k+1$ transitions in \eqref{eq:probtoyfk}. 
We will then go on to point out that there is a fundamental 
structural reason for the fact that the attempts to interpret the 
full model evolution given in \eqref{eq:modbjimwlk}
in a probabilistic manner have run into problems. We will see that 
it is not possible to interpret this equation using a probabilistic 
$2 \to 1$ vertex even in the toy model. However, we note that 
it is not necessary to include a $2 \to 1$ vertex to obtain saturation. 
In fact any $2\to n$ vertex where only one of the $n$ dipoles interact 
also corresponds to a $2\to 1$ transition, and this will be discussed 
more later.

First we write the $S$-matrix given in \eqref{eq:toySmatrix} as 
\begin{eqnarray}
\mathcal{S}(Y_1+Y_2)=\pmb{p}^T(Y_1)\pmb{s}_{\pmb{q}}(Y_2), \, \, \, \,
\pmb{s}_{\pmb{q}} \equiv \pmb{S}\pmb{q}
\end{eqnarray} 
where the $n$th component of the vector $\pmb{s}_{\pmb{q}}(Y_2)$ is the $S$-matrix
element of a projectile, evolved up to $Y_1$, made up from $n$ dipoles scattering against 
a generic target, which is evolved up to $Y_2$. It is then easy to see that $(\pmb{s}_{\pmb{q}})_n 
\equiv \langle s^n \rangle$ satisfies the following evolution equation \cite{Blaizot:2006wp}
\begin{eqnarray}
\partial_Y  \langle s^n \rangle = \mathcal{R}(n) \{ \langle s^{n+1} \rangle - \langle s^n \rangle \} 
\end{eqnarray}

Using the relation $s=1 - t$, where 
$t$ denotes the scattering amplitude, one can similarly derive the equations 
obeyed by $\langle t^n\rangle$. Since $\tau$ is assumed to be small, one 
can expand $\mathcal{R}(n)$ in each equation and drop contributions which are 
negligible in all regimes (dilute and dense systems). Doing this, the authors 
in \cite{Blaizot:2006wp} arrived at the following evolution equations for $\langle t\rangle$, 
$\langle t^2\rangle$ and $\langle t^3\rangle$,   
\begin{eqnarray}
\partial_Y \langle t\rangle &=& \langle t\rangle - \langle t^2\rangle, \label {eq:t1}\\
\partial_Y \langle t^2\rangle &=& 2(\langle t^2\rangle - \langle t^3\rangle)
+\tau \langle t(1-t)^2\rangle, \label{eq:t2} \\
\partial_Y \langle t^3\rangle &=& 3(\langle t^3\rangle - \langle t^4\rangle)
+ 3\tau \langle t^2(1-t)^2\rangle + \tau^2  \langle t(1-t)^3\rangle.
\label{eq:t3}
\end{eqnarray}
If one neglects all terms proportional to $\tau$, then it is seen that the 
resulting hierarchy corresponds to the large $N_c$ version of the B-JIMWLK 
hierarchy. 

Let us now see how equations \eqref{eq:t1}-\eqref{eq:t3} arise from the transition rates 
$\mathcal{R}_{k\to k+1}^{(n)}$. The evolution of the $S$-matrix elements are given by 
\begin{eqnarray}
\partial_Y  \langle s\rangle &=& \mathcal{R}^{(1)}_{1\rightarrow 2} 
\{\langle s^2\rangle - \langle s\rangle \}, \\
\partial_Y  \langle s^2\rangle &=& (\mathcal{R}^{(2)}_{1\rightarrow 2} + \mathcal{R}^{(2)}_{2\rightarrow 3}) 
\{\langle s^3\rangle - \langle s^2\rangle \}, \\
\partial_Y  \langle s^3\rangle &=& (\mathcal{R}^{(3)}_{1\rightarrow 2} + \mathcal{R}^{(3)}_{2\rightarrow 3}
+ \mathcal{R}^{(3)}_{3\rightarrow 4}) \{\langle s^4\rangle - \langle s^3\rangle \}.
\end{eqnarray}
It is then straightforward to obtain the following equations for the scattering
amplitudes 
\begin{eqnarray}
\partial_Y  \langle t\rangle &=& \mathcal{R}^{(1)}_{1\rightarrow 2}\{\langle t\rangle - 
\langle t^2\rangle \} = \langle t\rangle - \langle t^2\rangle, \\
\partial_Y  \langle t^2\rangle &=& 2\mathcal{R}^{(1)}_{1\rightarrow 2}(\langle t\rangle 
- \langle t^2\rangle) - \mathcal{R}^{(2)}_{1\rightarrow 2}\langle t(1 - t)^2 \rangle
- \mathcal{R}^{(2)}_{2\rightarrow 3}\langle t(1 - t)^2\rangle \nonumber \\
&=& 2(\langle t^2\rangle - \langle t^3\rangle) + 2\tau \langle t(1-t)^2\rangle
- \tau \langle t(1-t)^2\rangle. 
\label{eq:t2ea}
\end{eqnarray}
We indeed see that the first equation is equal to \eqref{eq:t1} and that 
the second equation is equal to \eqref{eq:t2}. It is also straightforward 
to see that the equation for $\langle t^3\rangle$ agrees with \eqref{eq:t3}. 

Next, we comment on the structure of the equation given 
in \eqref{eq:modbjimwlk}. Assume we wish to view the process as a stochastic 
evolution with a $1\to 2$ splitting vertex, $f_{1\to 2}$, and a $2\to 1$ merging 
vertex, $k_{2\to 1}$. We here assume the total splitting rate to be the incoherent sum 
of the individual splitting rates. For the evolution of the $2-$dipole state 
we have 
\begin{eqnarray}
\partial_Y \langle s^2\rangle = f_{1\to 2}^{(2)}\{\langle s^3\rangle - \langle s^2\rangle\}
+ k_{2\to 1}^{(2)}\{\langle s\rangle - \langle s^2\rangle\}
\end{eqnarray}
which gives 
\begin{eqnarray}
\partial_Y \langle t^2\rangle &=& 2f_{1\to 2}^{(1)}\{\langle t\rangle - \langle t^2\rangle\}
+  k_{2\to 1}^{(2)}\langle t(1-t)\rangle - f_{1\to 2}^{(2)}\langle t(1-t)^2\rangle \nonumber \\
&=& (2f_{1\to 2}^{(1)}-f_{1\to 2}^{(2)}+k_{2\to 1}^{(2)})\langle t\rangle + 
(-2f_{1\to 2}^{(1)} +2f_{1\to 2}^{(2)}-k_{2\to 1}^{(2)})\langle t^2\rangle
-  f_{1\to 2}^{(2)}\langle t^3\rangle \nonumber \\
&=& k_{2\to 1}^{(2)} \langle t\rangle + (f_{1\to 2}^{(2)} -k_{2\to 1}^{(2)})\langle t^2\rangle - 
f_{1\to 2}^{(2)}\langle t^3\rangle.
\end{eqnarray}
We thus see that the $2\to 1$ contribution not only generates the ``fluctuation'' term, 
$k_{2\to 1}^{(2)} \langle t\rangle$, but it 
does also modify the $\langle t^2\rangle$ term. If this term is to be unaffected by the additional vertex, as 
in \eqref{eq:modbjimwlk}, then we have to set $k_{2\to 1}^{(2)}=0$. This is actually 
very similar to what happens in the full model. In that case the integral over the proposed
$2\to 1$ vertex has to be zero, which implies that the vertex cannot be positive definite, 
as was noted in  \cite{Iancu:2005dx}. We can also try to add another 
vertex such that the total contribution to the $\langle t^2\rangle$ term cancels. Assume for example 
the existence 
of an additional $2\to 0$ vertex $g_{2\to 0}$. We then get 
\begin{eqnarray}
\partial_Y \langle t^2\rangle = (k_{2\to 1}^{(2)} + 2g_{2\to 0}^{(2)})\langle t\rangle + 
(f_{1\to 2}^{(2)} - k_{2\to 1}^{(2)} - g_{2\to 0}^{(2)})\langle t^2\rangle - f_{1\to 2}^{(2)} \langle t^3\rangle
\end{eqnarray} 
from which we conclude that 
\begin{eqnarray}
k_{2\to 1}^{(2)} + g_{2\to 0}^{(2)} = 0
\end{eqnarray}
which means that either $k_{2\to 1}^{(2)}$ or $g_{2\to 0}^{(2)}$ has to be negative. 
Thus we conclude that this approach has big problems, as one must choose 
$k_{2\to 1}^{(2)}$ either to be 0, or it must be negative, which means that one 
cannot obtain a probabilistic formulation.

\subsection{The 1+1 dimensional toy model}
\label{sec:onetoymod}

A somewhat more complicated 1+1 dimensional model is presented in \cite{Iancu:2006jw}.
The structure of this model is very similar to the 1+0 dimensional model, 
but the difference is that this time a dipole state is not only specified 
by the total number of dipoles, but it also depends on the distribution of these dipoles 
along some additional transverse axis. 

%This axis can assumed to be either discrete or continuous. 
%For a discrete axis (a lattice), a dipole state is  
%specified by giving the occupation number on the lattice sites, $n_i$, and 
%the general state is denoted $|\{n\}\rangle = |n_0, \dots, n_i, \dots \rangle$.  

%\begin{eqnarray}
%\pmb{\mathcal{H}}(\{n\} \rightarrow \{m\}) = \langle \{m\}| \pmb{\mathcal{H}}
%|\{n\}\rangle = \sum_i \mathcal{R}_i(\{n\})(\delta_{m_i,n_{i}+1} \prod_{j\neq i} \delta_{m_j,n_j}
%- \prod_{j} \delta_{m_j,n_j})
%\end{eqnarray}
%where $\mathcal{R}_i(\{n\})$ is the probability per unit rapidity to add an extra 
%dipole on site $i$ for the configuration $|\{n\}\rangle$. In a collision, a dipole at site 
%$i$ and a dipole at site $j$ scatter with a probability $\tau(i|j)\equiv \tau_{ij}$ 
%and the $S$-matrix element between the states $\langle \{m\}|$ and $|\{n\}\rangle$ 
%is given by $\pmb{\mathcal{S}}(\{n\},\{m\}) = \prod_{ij}(1-\tau_{ij})^{n_im_j}$. 

%\begin{eqnarray}
%\frac{\mathcal{R}_i(\{n\})}{\Delta} = \frac{1 - \prod_j (1 - \tau_{ij})^{n_j}}{\tau}.
%\label{eq:1dimfdisc}
%\end{eqnarray}
%Here $\Delta$ is the lattice spacing and the splitting rate $\mathcal{R}_i(\{n\})/\Delta$ is 
%well defined in the continuum limit. In the above equation 
%$\tau \equiv \tau_{ii}$ (no summation) is assumed to be independent of $i$. 
We denote the position of a dipole along this 
axis with $x_i$, and the generic $n$-dipole state is denoted  
$| \{x_i\}\rangle = |x_1, \dots, x_n \rangle$. The assumption in \cite{Iancu:2006jw} 
is that the dipole state evolves \emph{only} by 
the addition of a single new dipole at some position $x_{n+1}$. In that case the frame independence 
equation in \eqref{eq:boostinv} can easily be solved, and the simplest 
solution for the total splitting rate $\mathcal{R}_i(\{n\})$ 
is given by \cite{Iancu:2006jw}
\begin{eqnarray}
\mathcal{R}(\{x_i\}\rightarrow \{x_i\} + x_{n+1}) = \frac{1 - \prod_{i=1}^n (1 - \tau(x_i|x_{n+1}))}{\tau}.
\label{eq:1dimfcont}
\end{eqnarray}
Here $\tau$ is a constant which can, by a redefinition of $Y$, set to be equal to 1. 

%Note the similarity of \eqref{eq:1dimfdisc} and \eqref{eq:1dimfcont} to \eqref{eq:toyf}.
%In fact this should not come as a surprise since both models are very similar indeed. 
%Even though the one dimensional model contains an additional dimension, it is nevertheless
%assumed that the dipole state evolves by the addition of a new dipole 
%without changing the configuration of the state which produced this new dipole. 
%This assumption makes the two models quite similar.  

We now show that the evolution can once again be formulated 
as a probabilistic process in terms of coherent $k\to k+1$ transitions 
as in sec \ref{sec:stochmod}. In this case we assume we have a system 
of dipoles, $X_i$, which live on a one dimensional spatial axis. 
We assume this axis to represent the position of the ``point-like'' dipoles. 
This axis is assumed to be continuous, so that the index $i$ actually 
represents a continuous label $x_i$. An isolated dipole $X_i$ can then emit another dipole 
$X_j$ at position $x_j$ with a probability $\tau_{ij}=\tau(x_i|x_j)$. However, in the presence 
of more than one $X_i$, 
the new $X_j$ can also be emitted coherently from several dipoles with a probability given by 
the product of the individual emission probabilities. For a system of $n$ dipoles 
located at positions $x_1,\dots, x_n$, the total $k\to k+1$ splitting 
rates $\mathcal{R}_{k\to k+1}^{(n)}(\{x_i\}\to \{x_i\} + x_{n+1})$ are then given by 
\begin{eqnarray}
\mathcal{R}_{k\to k+1}^{(n)}(\{x_i\}\to \{x_i\} + x_{n+1}) = 
\frac{1}{k!}\sum_{i_1 \neq \dots \neq i_k}^n \tau_{i_1,n+1}
\tau_{i_2,n+1}\dots \tau_{i_k,n+1} \prod_{m\neq 1, \dots, k}^n (1-\tau_{i_m,n+1}), \nonumber \\
\label{eq:fkonedim}
\end{eqnarray} 
and their sum satisfies
\begin{eqnarray}
\sum_{k=1}^n \mathcal{R}_{k\to k+1}^{(n)} (\{x_i\}\to \{x_i\} + x_{n+1})= 1 - \prod_{i=1}^n
(1 - \tau_{i, n+1}).
\label{eq:fprobsum}
\end{eqnarray}
which is equal to \eqref{eq:1dimfcont}. 
Note that once again the positive definite splitting rates do not only depend on the 
state of the emitting dipoles, but also on the state of the other dipoles in the 
cascade. This is unavoidable if one wants to obtain a probabilistic 
evolution. 

The $k\to k+1$ splitting rates in \eqref{eq:fkonedim} are very similar in structure  
to the processes generated by the dipole swing, to be discussed in the forthcoming sections. 
Anticipating the discussion there, we can interpret \eqref{eq:fkonedim} as a process 
where the newly produced dipoles swing multiply with the rest of the dipoles in the cascade
(a concrete example of this is shown in fig \ref{fig:3pomtopology}). Note that in the 
toy models both the splitting and the scatterings of the dipoles are determined by the 
quantities $\tau_{ij}$. If we would assume that these 
also determine the swing probability, then the $k\to k+1$ splitting rates in \eqref{eq:fkonedim} 
would describe processes where the newly produced dipole $i$ swings with $k-1$ dipoles 
from the cascade, and the factor $\prod_j(1-\tau_{ij})$ could then be interpreted as the 
probability that $i$ swings with no more than $k-1$ dipoles. 

One difficulty is, however, that the swing in its 
form in the full model cannot really provide saturation in the toy models 
since the dipoles have no size here \footnote{In \cite{Iancu:2006jw}, the spatial 
axis was interpreted as being related to the dipole size while we feel a more 
close analogy is to interpret it as a spatial coordinate where dipoles of 
some fixed size live. Irrespective of the interpretation, however, direct 
comparison with the full model is made difficult by the 
assumption that the toy model state only evolves by the addition of a single dipole.}. 
This follows from the fact that both toy models have trivial topologies, 
in the sense that the dipole state is assumed to evolve only by the addition of a new 
dipole without changing the emitting state.
In the toy models saturation 
occurs because $k$ dipoles emit a single dipole with the same strength 
as a single dipole. In our implementation of the dipole swing however, 
saturation occurs because the swing decreases the sizes of the dipoles, and 
smaller dipoles have a smaller probability to interact.

\section{Evolution in the full model}
\label{sec:fulldipmod}

We will here argue that the evolution in the full model 
also can be formulated as a probabilistic process in terms of 
$k\to k+1$ transitions.

%First we recall that the $S$-matrix is in the full model given by 
%\begin{eqnarray}
%S_Y &=& \sum_{N,M} P_N(Y_0)P_M(Y-Y_0)\,\mathrm{exp}\biggl ( -\sum_{i=1}^N\sum_{j=1}^M
%f(x_i,y_i|x_j,y_j)\biggr ) \nonumber \\
%&=& \biggl \langle \mathrm{exp}\biggl ( -\sum_{i,j}
%f_{ij} \biggr ) \biggr \rangle.
%\label{eq:Smatrix}
%\end{eqnarray}
%Note that each $f_{ij}$ is of order $\alpha_s^2$, and in order to make 
%the analogy to the 1+1 dimensional toy model closer, we rewrite each factor $e^{-f_{ij}}$
%as $1-f_{ij}$. Thus we have 
The $S$-matrix is in the full model given by 
\begin{eqnarray}
S_Y &=& \sum_{N,M} P_N(Y_0)P_M(Y-Y_0)\prod_{i=1}^N\prod_{j=1}^M
\biggl (1 - f(x_i,y_i|x_j,y_j) \biggr). \nonumber \\
&=& \biggl \langle \prod_{i=1}^N\prod_{j=1}^M  (1 - f_{ij} ) \biggr \rangle.
\label{eq:Smatrix}
\end{eqnarray}
where $f_{ij}$ is given by \eqref{eq:dipamp}.

Let us consider the evolution initiated by a pair of oppositely moving $q\bar{q}$
pairs.
We then consider generic dipole states $\mathscr{A}_N$, containing $N-1$ gluons, 
which at each rapidity step can evolve into states $\mathscr{A}_{N+1}$ containing 
$N$ gluons. The splitting rate is denoted $\mathcal{R}(\mathscr{A}_N \to \mathscr{A}_{N+1})$. 

The scattering between the states  $\mathscr{A}_N$ and $\mathscr{B}_M$ is then 
frame independent if
\begin{eqnarray}
\sum_{\mathscr{A}_{N+1}} \int_{z} \mathcal{R}(\mathscr{A}_N \to \mathscr{A}_{N+1}) \biggl \{ 
1 - \prod_{j\in \mathscr{B}_M}\{ 1 - (\sum_{i \in \mathscr{A}_{N+1}\setminus 
\mathscr{A}_N} - \sum_{i \in \mathscr{A}_{N}\setminus\mathscr{A}_{N+1}})f(i|j)\}\biggr \}
= \nonumber \\
\sum_{\mathscr{B}_{M+1}} \int_{z} \mathcal{R}(\mathscr{B}_M \to \mathscr{B}_{M+1}) \biggl \{ 
1 - \prod_{i\in \mathscr{A}_N}\{ 1 - (\sum_{j \in \mathscr{B}_{M+1}\setminus 
\mathscr{B}_M} - \sum_{j \in \mathscr{B}_{M}\setminus \mathscr{B}_{M+1}})f(i|j)\}\biggr \}.
\nonumber \\
\label{eq:frameindp}
\end{eqnarray} 
The notations in this equation are as follows. The integral $\int_{z}$ denotes the 
integration over the transverse position of the $N$th emitted gluon. 
In $\prod_{j\in \mathscr{B}_M}$, the 
index $j$ runs over all dipoles in the state $\mathscr{B}_M$. The set denoted 
by $\mathscr{A}_{N+1}\setminus \mathscr{A}_N$ consists of those new dipoles 
produced in the last step of the evolution. Similarly, $\mathscr{A}_{N}\setminus \mathscr{A}_{N+1}$
denotes the set of all dipoles which are present in $\mathscr{A}_{N}$, but not in 
$\mathscr{A}_{N+1}$, \emph{i.e.} those dipoles which were removed from the cascade 
in the last step of the evolution. Finally, $f(i|j)$ stands for the scattering 
amplitude between the dipoles $i$ and $j$, \emph{i.e.} the expression $f_{ij}$ in 
\eqref{eq:dipamp} and \eqref{eq:Smatrix}. The sum $\sum_{\mathscr{A}_{N+1}}$ is over all $N$ gluon 
states which can be reached from $\mathscr{A}_N$ in one step. 

In the toy model analogy, the difference $\sum_{i \in \mathscr{A}_{N+1}\setminus 
\mathscr{A}_N} - \sum_{i \in \mathscr{A}_{N}\setminus\mathscr{A}_{N+1}}$ consists of only
the newly produced dipole, $i$, since all others dipoles are assumed to be 
unaffected by the evolution. In that case eq \eqref{eq:frameindp} reduces to 
\begin{eqnarray}
\sum_i \mathcal{R}(\mathscr{A}_N \to \mathscr{A}_N + i) \biggl \{ 
1 - \prod_{j\in \mathscr{B}_M} ( 1 - f(i|j)) \biggr \} \nonumber \\
= \sum_i \mathcal{R}(\mathscr{B}_M \to \mathscr{B}_M + i) \biggl \{ 
1 - \prod_{j\in \mathscr{A}_M} ( 1 - f(i|j)) \biggr \}.
\end{eqnarray}
The most simple solution is given by
\begin{eqnarray}
\mathcal{R}(\mathscr{A}_N \to \mathscr{A}_N + i) = 1 - \prod_{j\in \mathscr{A}_N} ( 1 - f(i|j)),
\end{eqnarray}
which we recognize from \eqref{eq:1dimfcont}. We will now see that the 
evolution in the full model can be formulated probabilistically in terms of 
$k\to k+1$ vertices as in the toy models. At first we will formulate 
the evolution as in eq \eqref{eq:toyfk} which implies that the $2\to 3$ transition 
will appear to have negative sign. However, we know from above how to 
treat these signs, and thus give the evolution a probabilistic interpretation.
We will also see how one can interpret these vertices in terms of the dipole swing. 

To this end, we consider first the situation where $N=2$ and $M=1$. The state $\mathscr{A}_2$ 
must consist of two connected dipoles since we know that a single, isolated dipole (the 
state $\mathscr{A}_1$) 
evolves by a dipole splitting (assuming the evolution is initiated by a single $q\bar{q}$
pair). We then denote the two dipoles 
in $\mathscr{A}_2$ with $a$ and $b$, and the single dipole in the state $\mathscr{B}_1$
is denoted by $l$. We can then write \eqref{eq:frameindp} as
\begin{eqnarray}
\sum_{\mathscr{A}_3} \int_{z} \mathcal{R}(a,b \to \mathscr{A}_3) ( 
\sum_{i \in \mathscr{A}_{3}\setminus 
\mathscr{A}_2} - \sum_{i \in \mathscr{A}_{2}\setminus\mathscr{A}_{3}} ) f(i|l) = 
\int_{z} \mathcal{R}(l\to (l_1,z)+(z,l_2)) \times \nonumber \\
\times \biggl ( \mathscr{F}(l, a, z) + \mathscr{F}(l, b, z) - \mathscr{F}(l, a, z)\mathscr{F}(l, b, z)
\biggr ) 
\label{eq:frameindp2}
\end{eqnarray} 
where 
\begin{eqnarray}
\mathscr{F}(l, a, z) = f(l_1, z|a) + f(z, l_2|a) - f(l|a).
\end{eqnarray}
Here $l_1$ and $l_2$ denote the transverse positions of the partons 
of dipole $l$. From studying the case $N=M=1$ (the scattering between two elementary dipoles), 
we know that each isolated dipole 
evolves by a $1\to 2$ splitting. We therefore write $\sum_{\mathscr{A}_3} 
\mathcal{R}(a,b \to \mathscr{A}_3)$ as 
\begin{eqnarray}
\sum_{\mathscr{A}_3} 
\mathcal{R}(a,b \to \mathscr{A}_3) &=& \mathcal{R}^{(1)}(a \to 
(a_1, z) + (z, a_2)) + \mathcal{R}^{(1)}(b \to (b_1, z) + (z, b_2)) \nonumber \\ 
&+& \sum_{\mathscr{A}_3^{(2)}} \mathcal{R}^{(2)}(a, b \to \mathscr{A}_3^{(2)}).
\end{eqnarray}
Here $\mathscr{A}_3^{(2)}$ denotes the set of all 2 gluon states which can be reached 
from $\mathscr{A}_2=\{a,b\}$ via the vertex $\mathcal{R}^{(2)}$.
Now, we know from \cite{Mueller:1996te} that the incoherent contributions, $\mathcal{R}^{(1)}$,
above are equal to the first order
contributions (in $f(i|j)$) in \eqref{eq:frameindp2}. Thus we are left with the equation 
\begin{eqnarray}
\sum_{\mathscr{A}_3^{(2)}} \int_{z} \mathcal{R}^{(2)}(a,b \to \mathscr{A}_3^{(2)}) ( 
\sum_{i \in \mathscr{A}_{3}^{(2)}\setminus 
\mathscr{A}_2} - \sum_{i \in \mathscr{A}_{2}\setminus\mathscr{A}_{3}^{(2)}} ) f(i|l) = \nonumber \\
= - \int_{z} \mathcal{R}(l\to (l_1,z)+(z,l_2)) 
\biggl (\mathscr{F}(l, a, z) \cdot \mathscr{F}(l, b, z)
%f(l_1, z|a) + 
%f(z, l_2|a) - f(l|a))\times (f(l_1, z|b) + f(z, l_2|b) - f(l|b)
\biggr ). 
\label{eq:frameindp3}
\end{eqnarray}
In the toy model, where the dipole state evolves by the addition of a single dipole 
$i$ only, we know that 
$\mathcal{R}^{(2)}(a,b \to \mathscr{A}_3^{(2)}) =  -f(i|a)f(i|b)$ (in case 
we use the formulation in \eqref{eq:toyfk}). 
%\footnote{Here we have 
%included a minus sign which would not have appeared if we had multiplied the incoherent 
%transition rates by a factor similar to $1-\tau$ as in the toy model. Here we are not 
%interested in the overall sign of this vertex, but only on what transition it describes.}. 
Indeed in that case 
we see that both sides in \eqref{eq:frameindp3} equals $-\sum_i f(i|a)f(i|b)
f(i|l)$. It is then clear that we must have a $2\to 3$ transition. 

%In order for the full model equation to agree with the results from the toy model, 
%we must clearly have $\biggl | \sum_{i \in \mathscr{A}_{3}^{(2)}\setminus 
%\mathscr{A}_2}\biggr | - \biggl |\sum_{i \in \mathscr{A}_{2}\setminus\mathscr{A}_{3}^{(2)}} \biggr| = 1$,
%where $\biggl |\sum \biggr |$ denotes the number of terms in the sum. 
%Since we only have two incoming dipoles, $a$ and $b$, the only possibility is 
%that we have 
%$\biggl |\sum_{i \in \mathscr{A}_{3}^{(2)}\setminus\mathscr{A}_{2}} \biggr| = 3$. 
%Thus we must have a $2\to 3$ transition. 

By similarly studying the case where $N=3$ and $M=1$, 
we would conclude that we need an 
additional $3\to 4$ dipole vertex and so on. We are thus led to a picture where the dipole 
state evolves by $k \to k+1$ transitions. If these transitions can be generated by combining the 
dipole splitting with the dipole swing, as we will argue in the next sections, 
we furthermore obtain a probabilistic interpretation of the evolution, as 
was discussed in the end of sec \ref{sec:onetoymod}. 
%Of course this does not prove the existence 
%of such vertices, which solve eq \eqref{eq:frameindp}. It does, however, show that 
%this is the only possibility for the evolution of the dipole state. 

Before we going on, we also note that 
care has to be taken to the fact that the frame independence equation 
in \eqref{eq:frameindp} may contain divergences. In the original 
dipole model, these divergences arise from the dipole splitting kernel 
in \eqref{eq:dipkernel}, but the frame independence equation is still 
finite since the expressions in the brackets in \eqref{eq:frameindp} 
vanish at these singular points. This is both due to the topology of the 
dipole splitting and also to the colour transparency of small dipoles. 
Any new vertex to be introduced into the model must retain this property 
since otherwise equation \eqref{eq:frameindp} would not make any sense. 
 
\section{The Colour Topology of the Evolution}
\label{sec:evtopology}

\subsection{Colour Flow}

Although we are not able to explicitely write down the  
splitting rate $\mathcal{R}^{(3)}(a,b \to \mathscr{A}_3^{(2)})$ in the full model, 
we will in this section argue that the correct topology of the evolution 
is the one induced by the dipole swing. 
%Furthermore the dipole swing 
%naturally provides us with $k \to k+1$ transitions in the evolution.

We first write eq \eqref{eq:frameindp} in Mueller's original formulation, 
\begin{eqnarray}
\sum_{i=1}^N \int_{z} \mathcal{M}(i|z) \biggl \{
1 - \prod_{j=1}^M(1 - (\sum_{k\in new} - \sum_{k \in old})f(k|j)) \biggr \} = \nonumber \\
= \sum_{i=1}^M \int_{z} \mathcal{M}(i|z) \biggl \{
1 - \prod_{j=1}^N(1 - (\sum_{k\in new} - \sum_{k \in old})f(k|j)) \biggr \}.
\label{eq:Mulframeindp}
\end{eqnarray}
Here $\mathcal{M}(i|z)$ is the usual dipole kernel in \eqref{eq:dipkernel}
for a dipole $i$ emitting 
a gluon at position $z$, and for simplicity we denote with $\sum_{k\in new (old)}$
the sum over the dipoles produced (destroyed) in the last step. 

\FIGURE[t]{
\scalebox{0.9}{\mbox{
\begin{picture}(200,230)(20,40)
\Vertex(10,230){2}
\Text(5,230)[r]{$x_1$}
\Vertex(10,200){2}
\Text(5,200)[r]{$y_1$}
\ArrowLine(10,230)(10,200)
\Gluon(15,215)(38,222){2}{4}
\Vertex(50,230){2}
\Text(55,230)[l]{$v_1$}
\Vertex(50,200){2}
\Text(55,200)[l]{$u_1$}
\DashLine(50,200)(50,230){2}
\Vertex(35,215){2}
\Text(31,211)[r]{$z$}
\ArrowLine(35,215)(50,230)
\ArrowLine(50,200)(35,215)
\Vertex(10,180){2}
\Text(5,180)[r]{$x_2$}
\Vertex(10,150){2}
\Text(5,150)[r]{$y_2$}
\ArrowLine(10,180)(10,150)
\Gluon(15,165)(45,165){2}{4}
\Vertex(50,180){2}
\Text(55,180)[l]{$v_2$}
\Vertex(50,150){2}
\Text(55,150)[l]{$u_2$}
\ArrowLine(50,150)(50,180)
\LongArrow(75,190)(100,190)

\Vertex(130,230){2}
\Text(125,230)[r]{$x_1$}
\Vertex(130,200){2}
\Text(125,200)[r]{$y_1$}
\ArrowLine(130,230)(170,230)
\Vertex(170,230){2}
\Text(175,230)[l]{$v_1$}
\Vertex(170,200){2}
\Text(175,200)[l]{$u_1$}
\Vertex(155,215){2}
\Text(157,208)[r]{$z$}
\ArrowLine(155,215)(130,200)
\ArrowLine(170,200)(155,215)
\Vertex(130,180){2}
\Text(125,180)[r]{$x_2$}
\Vertex(130,150){2}
\Text(125,150)[r]{$y_2$}
\ArrowLine(130,180)(170,180)
\Vertex(170,180){2}
\Text(175,180)[l]{$v_2$}
\Vertex(170,150){2}
\Text(175,150)[l]{$u_2$}
\ArrowLine(170,150)(130,150)

\Vertex(10,100){2}
\Text(5,100)[r]{$x_1$}
\Vertex(10,70){2}
\Text(5,70)[r]{$y_1$}
\DashLine(10,100)(10,70){2}
\Vertex(20,85){2}
\Text(22,78)[]{$z$}
\ArrowLine(10,100)(20,85)
\ArrowLine(20,85)(10,70)
\Gluon(20,93)(45,85){2}{4}
\Vertex(50,100){2}
\Text(55,100)[l]{$v_1$}
\Vertex(50,70){2}
\Text(55,70)[l]{$u_1$}
\ArrowLine(50,70)(50,100)
\Vertex(10,50){2}
\Text(5,50)[r]{$x_2$}
\Vertex(10,20){2}
\Text(5,20)[r]{$y_2$}
\ArrowLine(10,50)(10,20)
\Gluon(15,35)(45,35){2}{4}
\Vertex(50,50){2}
\Text(55,50)[l]{$v_2$}
\Vertex(50,20){2}
\Text(55,20)[l]{$u_2$}
\ArrowLine(50,20)(50,50)
\LongArrow(75,60)(100,60)
\Text(85,0)[]{$(A)$}

\Vertex(130,100){2}
\Text(125,100)[r]{$x_1$}
\Vertex(130,70){2}
\Text(125,70)[r]{$y_1$}
\ArrowLine(130,100)(170,100)
\Vertex(170,100){2}
\Text(175,100)[l]{$v_1$}
\Vertex(170,70){2}
\Text(175,70)[l]{$u_1$}
\Vertex(140,85){2}
\Text(145,78)[r]{$z$}
\ArrowLine(140,85)(130,70)
\ArrowLine(170,70)(140,85)
\Vertex(130,50){2}
\Text(125,50)[r]{$x_2$}
\Vertex(130,20){2}
\Text(125,20)[r]{$y_2$}
\ArrowLine(130,50)(170,50)
\Vertex(170,50){2}
\Text(175,50)[l]{$v_2$}
\Vertex(170,20){2}
\Text(175,20)[l]{$u_2$}
\ArrowLine(170,20)(130,20)
\end{picture}
}}

\epsfig{file=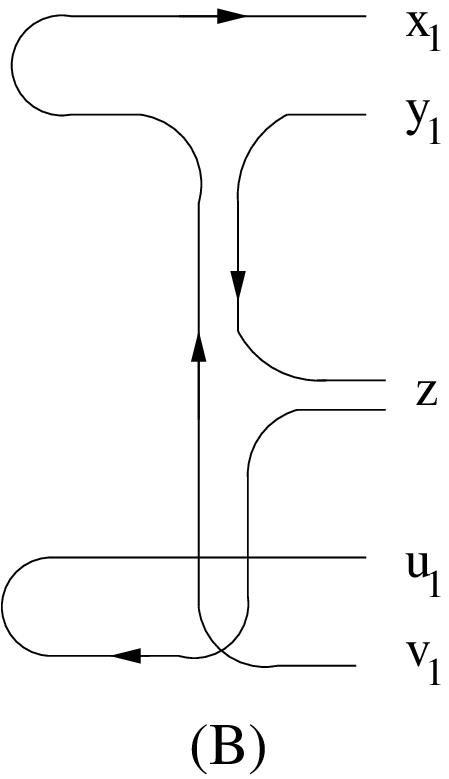,width=3cm}
\\
\\
\\
\caption
{\label{fig:1pomtopology} A process where the newly produced dipoles exchange 
   only a single gluon with the target. In this case the colour correlations 
  can be generated by the dipole splitting only. Here the position of $z$ 
  is integrated over and is therefore not fixed, and we only show one possible 
  colour flow. Note that each interaction 
  implies a change in the colour flow, which goes from colour to anticolour 
  as indicated by the arrows. It is not important 
  whether there are also other interactions or not. In fig $(A)$ we assume an 
  additional interaction between $(x_2,y_2)$ and $(u_2,v_2)$. In fig $(B)$ 
  we show the same colour flow in the corresponding Feynman diagram.}

}

In case the newly 
produced dipoles only scatter against one target dipole, eq \eqref{eq:Mulframeindp} 
linearizes, and in that case the equality is known to hold \cite{Mueller:1996te}. Note that this case 
does not restrict us to one pomeron exchange, it is only the dipoles produced 
in the last step which should scatter against a single dipole. There
may be still be several scatterings between dipoles produced earlier. 
In such events, the colour topology may be described by open chains 
stretching between the target and the projectile. An example is shown 
in fig \ref{fig:1pomtopology}. 
Here, the simple $1\to 2$ splitting is sufficient for producing 
all possible colour configurations.

\FIGURE[t]{
  \epsfig{file=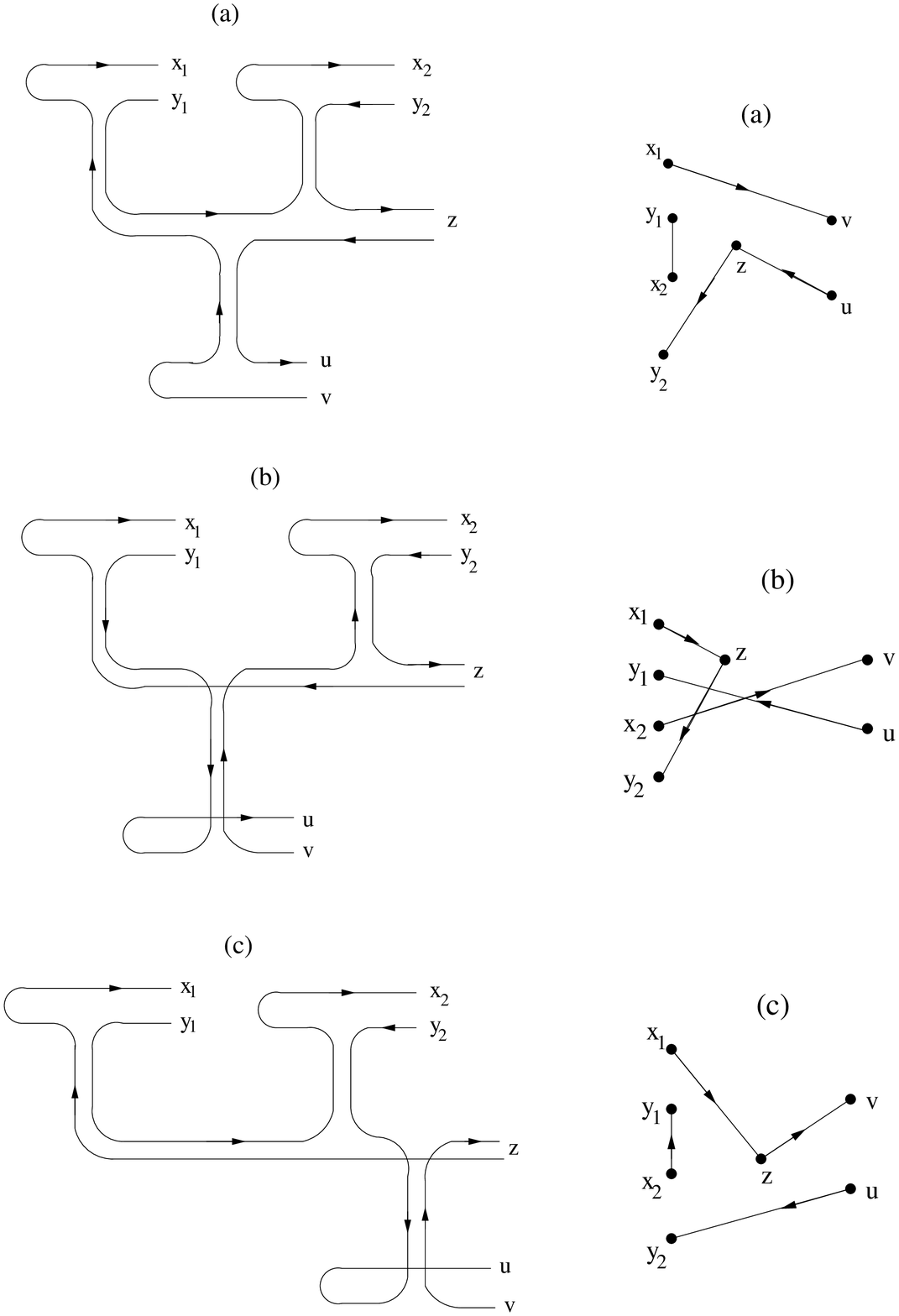,width=11cm}
 \caption{\label{fig:colorlines} The colour flow in a process where 
   two right-moving dipoles $(x_1,y_1)$ and $(x_2,y_2)$ are linked to 
   a single left-moving dipole $(u,v)$. 
   Three different configurations, $(a), (b)$ and $(c)$,  
   are shown. 
 }}

In case the newly produced dipoles scatter off two or more target dipoles, 
however, more complicated topologies are formed, and the simple $1\to 2$ splitting 
is not sufficient anymore. Let us first consider the case where each dipole 
is restricted to single scattering only. In this case, the newly produced 
two dipoles can at most scatter off two target dipoles. An example is shown in 
fig \ref{fig:colorlines} where three different colour configurations are formed (there are three 
more configurations which can be formed by reversing the colour flow in each line).
Here two right-moving dipoles $(x_1,y_1)$ and $(x_2, y_2)$, are connected to a single 
left-moving dipole $(u,v)$. With the restriction that each dipole scatters only once, 
the dipole $(u,v)$ can obviously not be connected to more than two oppositely 
moving dipoles. In this case one can produce all colour configurations by combining 
the dipole splitting with only one swing. This is in agreement with 
the findings in \cite{Kozlov:2006cg}. In fig \ref{fig:2pomtopology}
we show how the three configurations in fig \ref{fig:colorlines} can be generated 
from a dipole splitting and one swing, by first forming the configuration 
in fig \ref{fig:2pomtopology}$(A)$.
Figure \ref{fig:2pomtopology} can be compared to eq 
\eqref{eq:frameindp3} where the right hand side of that equation describes the evolution 
and the scattering of the dipole $(u,v)$ ($l$ in \eqref{eq:frameindp3}). The vertex 
$\mathcal{R}^{(2)}$ on the left hand side would then correspond to the diagrams showing 
the evolution of $(x_1, y_1)$ and $(x_2,y_2)$ into the three dipoles $(x_1,z), (z,y_2)$ 
and $(x_2,y_1)$. 

\FIGURE[t]{
\scalebox{0.75}{\mbox{
\begin{picture}(200,100)(50,-150)
\Vertex(10,100){2}
\Text(0,100)[]{$x_1$}
\Vertex(10,70){2}
\Text(0,70)[]{$y_1$}
\ArrowLine(10,100)(10,70)
\Vertex(10,30){2}
\Text(0,30)[]{$x_2$}
\Vertex(10,0){2}
\Text(0,0)[]{$y_2$}
\ArrowLine(10,30)(10,0)

\LongArrow(20,50)(40,50)

\Vertex(70,100){2}
\Text(60,100)[]{$x_1$}
\Vertex(70,70){2}
\Text(60,70)[]{$y_1$}
\DashLine(70,100)(70,70){3}
\ArrowLine(70,100)(100,60)
\ArrowLine(100,60)(70,70)
\Vertex(100,60){2}
\Text(102,55)[]{$z$}
\Vertex(70,30){2}
\Text(60,30)[]{$x_2$}
\Vertex(70,0){2}
\Text(60,0)[]{$y_2$}
\ArrowLine(70,30)(70,0)

\LongArrow(110,50)(130,50)

\Vertex(160,100){2}
\Text(150,100)[]{$x_1$}
\Vertex(160,70){2}
\Text(150,70)[]{$y_1$}
\DashLine(190,60)(160,70){3}
\ArrowLine(160,100)(190,60)
\Vertex(190,60){2}
\Text(192,55)[]{$z$}
\Vertex(160,30){2}
\ArrowLine(160,30)(160,70)
\Text(150,30)[]{$x_2$}
\Vertex(160,0){2}
\Text(150,0)[]{$y_2$}
\ArrowLine(190,60)(160,0)
\DashLine(160,30)(160,0){3}
\Text(100,-50)[]{$\large{(A)}$}

\end{picture}
}}
%\caption
%{\label{fig:2pomtopology} The configuration $\{(x_1,z), (z,y_2), (x_2,y_1)\}$ can 
%  be generated from $\{(x_1,y_1), (x_2,y_2)\}$ by combining a splitting and a swing.}

\scalebox{0.75}{\mbox{
\begin{picture}(200,400)(-50,0)
\Vertex(5,400){2}
\Text(0,400)[r]{$x_1$}
\Vertex(5,370){2}
\Text(0,370)[r]{$y_1$}
\ArrowLine(5,400)(30,360)
\ArrowLine(30,360)(5,300)
\Vertex(30,360){2}
\Text(32,355)[]{$z$}
\Vertex(5,330){2}
\ArrowLine(5,330)(5,370)
\Text(0,330)[r]{$x_2$}
\Vertex(5,300){2}
\Text(0,300)[r]{$y_2$}
\Vertex(70,340){2}
\Text(75,340)[l]{$u$}
\Vertex(70,370){2}
\Text(75,370)[l]{$v$}
\ArrowLine(70,340)(70,370)
\Gluon(25,380)(65,355){2}{5}

%\Text(250,350)[]{$\large{(B)}$}
\LongArrow(85,355)(105,355)

\Vertex(120,400){2}
\Text(115,400)[r]{$x_1$}
\Vertex(120,370){2}
\Text(115,370)[r]{$y_1$}
\DashLine(120,400)(145,360){3}
\ArrowLine(145,360)(120,300)
\Vertex(145,360){2}
\Text(147,355)[]{$z$}
\Vertex(120,330){2}
\ArrowLine(120,330)(120,370)
\Text(115,330)[r]{$x_2$}
\Vertex(120,300){2}
\Text(115,300)[r]{$y_2$}
\Vertex(185,340){2}
\Text(190,340)[l]{$u$}
\Vertex(185,370){2}
\Text(190,370)[l]{$v$}
\DashLine(185,340)(185,370){3}
\ArrowLine(120,400)(185,370)
\ArrowLine(185,340)(145,360)

\Vertex(5,250){2}
\Text(0,250)[r]{$x_1$}
\Vertex(5,220){2}
\Text(0,220)[r]{$y_1$}
\ArrowLine(5,250)(30,210)
\ArrowLine(30,210)(5,150)
\Vertex(30,210){2}
\Text(32,205)[]{$z$}
\Vertex(5,180){2}
\ArrowLine(5,180)(5,220)
\Text(0,180)[r]{$x_2$}
\Vertex(5,150){2}
\Text(0,150)[r]{$y_2$}
\Vertex(70,190){2}
\Text(75,190)[l]{$u$}
\Vertex(70,220){2}
\Text(75,220)[l]{$v$}
\ArrowLine(70,190)(70,220)
\Gluon(10,195)(65,195){2}{5}

%\Text(250,200)[]{\large{$(C)$}}
\LongArrow(85,205)(105,205)

\Vertex(120,250){2}
\Text(115,250)[r]{$x_1$}
\Vertex(120,220){2}
\Text(115,220)[r]{$y_1$}
\ArrowLine(145,210)(120,150)
\ArrowLine(120,250)(145,210)
\Vertex(145,210){2}
\Text(147,220)[]{$z$}
\Vertex(120,180){2}
\DashLine(120,180)(120,220){3}
\Text(115,180)[r]{$x_2$}
\Vertex(120,150){2}
\Text(115,150)[r]{$y_2$}
\Vertex(175,190){2}
\Text(180,190)[l]{$u$}
\Vertex(175,220){2}
\Text(180,220)[l]{$v$}
\DashLine(175,190)(175,220){3}
\ArrowLine(120,180)(175,220)
\ArrowLine(175,190)(120,220)

\Vertex(5,100){2}
\Text(0,100)[r]{$x_1$}
\Vertex(5,70){2}
\Text(0,70)[r]{$y_1$}
\ArrowLine(5,100)(30,60)
\ArrowLine(30,60)(5,0)
\Vertex(30,60){2}
\Text(32,55)[]{$z$}
\Vertex(5,30){2}
\ArrowLine(5,30)(5,70)
\Text(0,30)[r]{$x_2$}
\Vertex(5,0){2}
\Text(0,0)[r]{$y_2$}
\Vertex(70,40){2}
\Text(75,40)[l]{$u$}
\Vertex(70,70){2}
\Text(75,70)[l]{$v$}
\ArrowLine(70,40)(70,70)
\Gluon(24,30)(65,55){2}{5}

\LongArrow(85,55)(105,55)
\Text(100,-30)[]{$\large{(B)}$}

\Vertex(125,100){2}
\Text(120,100)[r]{$x_1$}
\Vertex(125,70){2}
\Text(120,70)[r]{$y_1$}
\ArrowLine(125,100)(150,60)
\DashLine(150,60)(125,0){3}
\Vertex(150,60){2}
\Text(152,55)[]{$z$}
\Vertex(125,30){2}
\ArrowLine(125,30)(125,70)
\Text(120,30)[r]{$x_2$}
\Vertex(125,0){2}
\Text(120,0)[r]{$y_2$}
\Vertex(190,40){2}
\Text(195,40)[l]{$u$}
\Vertex(190,70){2}
\Text(195,70)[l]{$v$}
\DashLine(190,40)(190,70){3}
\ArrowLine(150,60)(190,70)
\ArrowLine(190,40)(125,0)

\end{picture}
}}
\\
\\
\caption{\label{fig:2pomtopology} The three configurations marked by $(a), (b)$ and $(c)$ in 
  fig \ref{fig:colorlines} can all be generated   
  when the dipole $(u,v)$ interacts with one of three dipoles from 
  the configuration shown in fig $(A)$. This is illustrated in fig $(B)$.
  As 
  illustrated, the configuration in fig $(A)$ 
  can in turn be generated via the dipole swing. }
}

We conclude that in an approximation where multiple scatterings are allowed, but with 
the restriction that \emph{each dipole scatters only once}, the maximal correlation 
induced between the dipoles is that between a pair, and such a correlation 
can be generated by a simple swing.  

Actually, in this approximation, explicit frame independence in zero transverse 
dimensions can be achieved 
by including a $2 \to 1$ vertex in addition to the usual $1\to 2$ splitting. 
This reflects the fact that the maximal correlation induced is that 
between a pair of dipoles. Note also that, in the situation described above, 
it is always only one out of the three dipoles produced via the combination
of the dipole splitting and the swing 
which interacts with the target. In that 
sense the swing corresponds to a $2\to 1$ transition. We can thus obtain 
an effective  $2\to 1$ transition without actually decreasing the number of dipoles.  

If a single dipole can \emph{scatter multiply}, one swing will not be enough. 
In this case one dipole can for example split into two new dipoles, 
and these two dipoles 
can then interact with more than two target dipoles, inducing higher 
order correlations. Before going on, we note that there is an ambiguity 
in the statement that one dipole scatters multiply. Since each scattering 
implies a recoupling of the colour flow, a dipole which interacts is 
replaced by a new dipole. What we rather mean here is that the partons 
of the dipole can exchange multiple gluons.
%Likewise, if we wish to have several swings 
%at each step, we have to decide in which order the different swings 
%are performed, and the resulting topology will depend on this order.

\FIGURE[t]{
  \epsfig{file=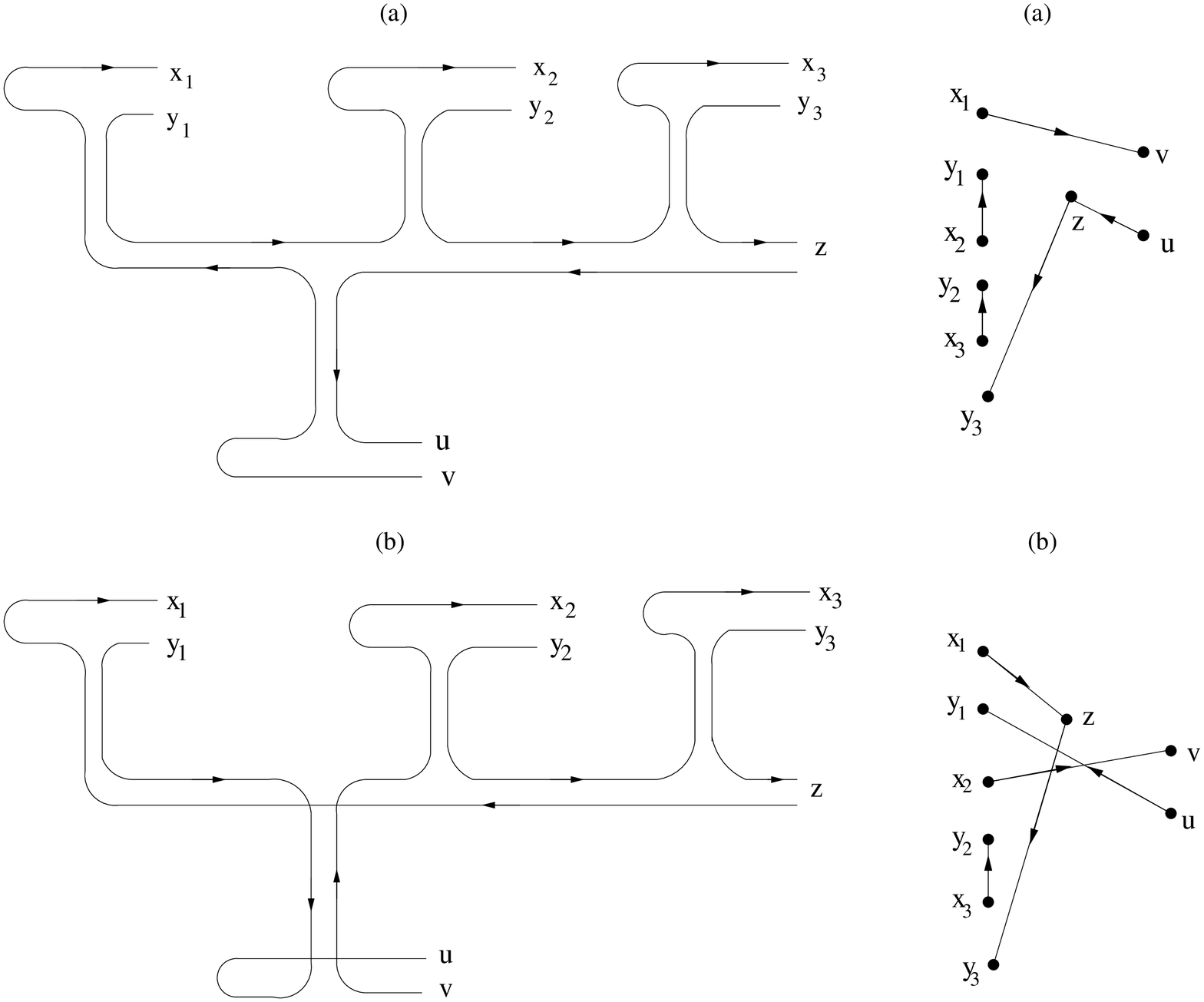,width=15cm}
 \caption{\label{fig:colorlines2} The colour flow in a process where 
   three right-moving dipoles $(x_1,y_1)$, $(x_2,y_2)$ and $(x_3,y_3)$ interact with 
   a single left-moving dipole $(u,v)$ after the emission of a 
   gluon located at $z$. For simplicity, we only show two, marked $(a)$ and $(b)$, 
   out of the four possible configurations. 
 }
}

\FIGURE{
\scalebox{0.7}{\mbox{
\begin{picture}(400,150)(-25,0)

\Vertex(10,150){2}
\Text(0,150)[]{$x_1$}
\Vertex(10,120){2}
\Text(0,120)[]{$y_1$}
\ArrowLine(10,150)(10,120)
\Vertex(15,90){2}
\Text(5,90)[]{$x_2$}
\Vertex(15,60){2}
\Text(5,60)[]{$y_2$}
\ArrowLine(15,90)(15,60)
\Vertex(5,30){2}
\Text(0,30)[r]{$x_3$}
\Vertex(5,0){2}
\Text(0,0)[r]{$y_3$}
\ArrowLine(5,30)(5,0)

\LongArrow(40,75)(70,75)

\Vertex(110,150){2}
\Text(100,150)[]{$x_1$}
\Vertex(110,120){2}
\Text(100,120)[]{$y_1$}
\DashLine(110,150)(110,120){3}
\Vertex(140,95){2}
\Text(142,90)[]{$z$}
\ArrowLine(110,150)(140,95)
\ArrowLine(140,95)(110,120)
\Vertex(115,90){2}
\Text(105,90)[]{$x_2$}
\Vertex(115,60){2}
\Text(105,60)[]{$y_2$}
\ArrowLine(115,90)(115,60)
\Vertex(105,30){2}
\Text(95,30)[]{$x_3$}
\Vertex(105,0){2}
\Text(95,0)[]{$y_3$}
\ArrowLine(105,30)(105,0)

\LongArrow(150,75)(180,75)

\Vertex(220,150){2}
\Text(210,150)[]{$x_1$}
\Vertex(220,120){2}
\Text(210,120)[]{$y_1$}
\Vertex(250,95){2}
\Text(252,90)[l]{$z$}
\ArrowLine(220,150)(250,95)
\DashLine(250,95)(220,120){3}
\ArrowLine(250,95)(225,60)
\Vertex(225,90){2}
\Text(215,90)[]{$x_2$}
\Vertex(225,60){2}
\Text(215,60)[]{$y_2$}
\DashLine(225,90)(225,60){3}
\ArrowLine(225,90)(220,120)
\Vertex(215,30){2}
\Text(205,30)[]{$x_3$}
\Vertex(215,0){2}
\Text(205,0)[]{$y_3$}
\ArrowLine(215,30)(215,0)

\LongArrow(270,75)(300,75)

\Vertex(340,150){2}
\Text(330,150)[]{$x_1$}
\Vertex(340,120){2}
\Text(330,120)[]{$y_1$}
\Vertex(370,95){2}
\Text(372,90)[l]{$z$}
\ArrowLine(340,150)(370,95)
\DashLine(370,95)(345,60){3}
\Vertex(345,90){2}
\Text(340,90)[r]{$x_2$}
\Vertex(345,60){2}
\Text(335,60)[]{$y_2$}
\ArrowLine(345,90)(340,120)
\Vertex(335,30){2}
\Text(325,30)[]{$x_3$}
\Vertex(335,0){2}
\Text(325,0)[]{$y_3$}
\DashLine(335,30)(335,0){3}
\ArrowLine(335,30)(345,60)
\ArrowLine(370,95)(335,0)
\Text(150,-20)[]{$\large{(A)}$}

\end{picture}
}}

\\
\\
\\
\\
\scalebox{0.7}{\mbox{
\begin{picture}(350,350)(-75,0)

\Vertex(10,350){2}
\Text(0,350)[]{$x_1$}
\Vertex(10,320){2}
\Text(0,320)[]{$y_1$}
\Vertex(40,295){2}
\Text(42,290)[l]{$z$}
\ArrowLine(10,350)(40,295)
\Vertex(15,290){2}
\Text(5,290)[]{$x_2$}
\Vertex(15,260){2}
\Text(5,260)[]{$y_2$}
\ArrowLine(15,290)(10,320)
\Vertex(5,230){2}
\Text(0,230)[r]{$x_3$}
\Vertex(5,200){2}
\Text(0,200)[r]{$y_3$}
\ArrowLine(40,295)(5,200)
\ArrowLine(5,230)(15,260)
\Vertex(80,300){2}
\Text(90,300)[]{$v$}
\Vertex(80,270){2}
\Text(90,270)[]{$u$}
\ArrowLine(80,270)(80,300)
\Gluon(26,330)(74,285){2}{6}
\Text(135,175)[r]{$\large{(B)}$}

\Text(135,-20)[]{$\large{(C)}$}

\Vertex(10,150){2}
\Text(0,150)[]{$x_1$}
\Vertex(10,120){2}
\Text(0,120)[]{$y_1$}
\ArrowLine(10,150)(10,120)
\Vertex(10,90){2}
\Text(0,90)[]{$x_2$}
\Vertex(10,60){2}
\Text(0,60)[]{$y_2$}
\ArrowLine(10,90)(10,60)
\Vertex(10,30){2}
\Text(0,30)[r]{$x_3$}
\Vertex(10,0){2}
\Text(0,0)[r]{$y_3$}
\ArrowLine(10,30)(10,0)
\Vertex(60,100){2}
\Text(70,100)[]{$v$}
\Vertex(65,70){2}
\Text(75,70)[]{$u$}
\Vertex(50,70){2}
\Text(50,65)[]{$z$}
\ArrowLine(65,70)(50,70)
\ArrowLine(50,70)(60,100)
\Gluon(53,97)(15,135){2}{6}
\Gluon(50,87)(15,75){2}{4}
\Gluon(47,77)(15,15){2}{6}

\Vertex(210,350){2}
\Text(200,350)[]{$x_1$}
\Vertex(210,320){2}
\Text(200,320)[]{$y_1$}
\Vertex(240,295){2}
\Text(242,290)[l]{$z$}
\ArrowLine(210,350)(240,295)
\Vertex(215,290){2}
\Text(205,290)[]{$x_2$}
\Vertex(215,260){2}
\Text(205,260)[]{$y_2$}
\ArrowLine(215,290)(210,320)
\Vertex(205,230){2}
\Text(200,230)[r]{$x_3$}
\Vertex(205,200){2}
\Text(200,200)[r]{$y_3$}
\ArrowLine(240,295)(205,200)
\ArrowLine(205,230)(215,260)
\Vertex(280,300){2}
\Text(290,300)[]{$v$}
\Vertex(280,270){2}
\Text(290,270)[]{$u$}
\ArrowLine(280,270)(280,300)
\Gluon(215,310)(274,289){2}{6}
%\Text(340,275)[r]{$\large{(C)}$}

%\Text(340,75)[]{$\large{(E)}$}
\Vertex(235,150){2}
\Text(225,150)[]{$x_1$}
\Vertex(235,120){2}
\Text(225,120)[]{$y_1$}
\ArrowLine(235,150)(235,120)
\Vertex(235,90){2}
\Text(225,90)[]{$x_2$}
\Vertex(235,60){2}
\Text(225,60)[]{$y_2$}
\ArrowLine(235,90)(235,60)
\Vertex(235,30){2}
\Text(225,30)[r]{$x_3$}
\Vertex(235,0){2}
\Text(225,0)[r]{$y_3$}
\ArrowLine(235,30)(235,0)
\Vertex(285,100){2}
\Text(295,100)[]{$v$}
\Vertex(280,55){2}
\Text(290,55)[]{$u$}
\Vertex(275,78){2}
\Text(280,78)[l]{$z$}
\ArrowLine(280,55)(275,78)
\ArrowLine(275,78)(285,100)
\Gluon(275,60)(240,135){2}{8}
\Gluon(275,91)(240,75){2}{4}
\Gluon(272,87)(240,15){2}{7}

\end{picture}

}}
\\
\caption{\label{fig:3pomtopology} The configurations marked by $(a)$ and $(b)$ in 
  figure \ref{fig:colorlines2} can be generated when the dipole $(u,v)$ interacts 
  with different dipoles from the configuration in fig $(A)$, as illustrated in fig
  $B$. There are two more configurations which can be obtained when $(u,v)$ interacts
  with the other two dipoles in fig $(A)$. In 
  fig $(C)$ we show how the same configurations can be generated when 
  the evolution is instead put into the dipole $(u,v)$.}
}

Consider the diagrams shown in fig \ref{fig:colorlines2}. Here a single 
left-moving dipole $(u,v)$ is linked to three right-moving dipoles
$(x_1,y_1), (x_2,y_2)$  and $(x_3,y_3)$, as shown in the figure. The 
two colour configurations shown in the figure can then be generated 
as illustrated in fig \ref{fig:3pomtopology}: First one generates 
the configuration $\{(x_3,y_2),(x_2,y_1), (x_1,z), (z,y_3)\}$ by 
combining a dipole splitting with two swings as shown in fig 
\ref{fig:3pomtopology}$(A)$ (this is obviously not the only process from which this 
final configuration can be generated). One of the four dipoles 
in this state can then collide with the dipole $(u,v)$. If for example $(u,v)$ 
collides with $(x_1,z)$, the configuration marked by $(a)$ in fig 
\ref{fig:colorlines2} is produced. 

The same process can also be viewed as an evolution of the dipole 
$(u,v)$, which then splits into $(u,z)$ and $(z,v)$,
and fig \ref{fig:3pomtopology} shows also how the two configurations in 
fig \ref{fig:colorlines2} can be generated when the dipoles $(u,z)$ and $(z,v)$
interact with $(x_1,y_1), (x_2,y_2)$ and $(x_3,y_3)$ exchanging now 3 
gluons (fig  \ref{fig:3pomtopology}$(C)$). 
Thus at least one of the 
dipoles $(u,z)$ and $(z,v)$ must scatter multiply in this case, since it would 
otherwise be impossible to generate the necessary colour correlations. As 
remarked in \cite{Blaizot:2006wp}, the evolution equations in sec \ref{sec:eveq}
actually describe such events where a newly produced dipole scatters 
multiply. We can also compare the processes in fig \ref{fig:3pomtopology} 
to eq \eqref{eq:frameindp} where one side of the equation  
describes the multiple scatterings of the dipoles $(u,z)$ and $(v,z)$, 
while the other side describes how the 3-dipole system evolves 
into a 4-dipole system which then exchanges a single gluon with $(u,v)$.

\subsection{Generating arbitrary correlations using at most $N-1$ swings}

We might then expect that the correlation induced by the scattering 
between a single left-moving dipole and $k$ right-moving dipoles
can be generated by a splitting followed by $k-1$ swings in the right-moving system. 
Note that such a process gives a $k\to k+1$ transition.
In the 
example above this was possible since we were able to form the 
configuration $\{(x_3,y_2),(x_2,y_1), (x_1,z), (z,y_3)\}$ by combining 
one splitting with two swings. This will always be possible if, given an 
arbitrary set of $N$ dipoles we always can generate 
all possible $(N+1)$-dipole 
states, by combining a splitting with at most 
$N-1$ dipole swings. We will below argue that this is indeed the case. 
In case we have $N$ spatially disconnected dipoles, the proof is 
easy. However, starting from a single $q\bar{q}$ pair, the evolution does not 
generate spatially disconnected dipoles, and in this case 
the result is a conjecture. 

\subsubsection{Spatially disconnected dipoles}
\label{sec:discdip}

As a warm up, we first show the statement in case we have $N$ spatially disconnected 
dipoles $\{(x_i,y_i)\}_{i=1}^N$. We then wish to evolve this state into some 
arbitrary $N+1$ dipole state, 
\begin{equation}
\prod_{i} (x_i, y_i) \to (x_k, z)
(z, y_j)\prod_{i \neq k, p(i) \neq j} (x_i, y_{p(i)}),
\end{equation}   
using at most $N-1$ dipole swings. Here $p(i)$ is a permutation of $i=1, \dots, N$. 
We first start by emitting gluon $z$ from the dipole $(x_k,y_k)$, 
\begin{equation}
\prod_{i} (x_i, y_i) \to (x_k,z)(z,y_k)\prod_{i\neq k} (x_i, y_i).
\end{equation}
The result then follows if we can show that, for an arbitrary permutation 
$p(i)$, we can with $N-1$ swings always make the transformation
\begin{equation}
\prod_{i=1}^N (x_i, y_i) \to \prod_{i=1}^N (x_i, y_{p(i)}).
\end{equation}
To this end, we perform the following swings in the indicated order, 
\begin{eqnarray}
&1& (x_N, y_N)(x_{p(N)}, y_{p(N)}) \to (x_N, y_{p(N)})(x_{p(N)}, y_{N}) \nonumber \\
&2& (x_{N-1}, y_{N-1})(x_{p(N-1)}, y_{p(N-1)}) \to (x_{N-1}, y_{p(N-1)})(x_{p(N-1)}, y_{N-1}) \nonumber 
\end{eqnarray}
etc. Then, after at most $N-2$ swings we are either finished, or we have 
\begin{equation}
(x_1, y_{p(2)})(x_2, y_{p(1)})\prod_{i=3}^N (x_i, y_{p(i)}).
\end{equation}
We then need only one more swing $(x_1, y_{p(2)})(x_2, y_{p(1)}) \to
(x_1, y_{p(1)})(x_2, y_{p(2)})$, and so after at most $N-1$ swings we are finished. 

The problem is that generally the dipoles are not spatially independent, and 
one then has to be careful in performing swings, since they might generate 
zero size dipoles, \emph{i.e.} colour singlet gluons which cannot be allowed.

\subsubsection{Dipole states initiated by a $q\bar{q}$ dipole}

\emph{Representation of the dipole states and the swing 
  in terms of permutations}
\\

Consider the evolution initiated by a $q\bar{q}$ colour dipole. In the 
original formulation of the dipole model, the dipole state at each rapidity 
$Y$ consists of an open chain, $\mathcal{C}$, of colour dipoles which are linked together via 
the gluons. Note the dual role played by the gluons 
and the dipoles,
each gluon links together two dipoles, and each dipole links together 
two gluons. 

The inclusion of the dipole swing generates closed dipole loops, 
$\mathcal{L}$, in addition to the open chain, $\mathcal{C}$. The swing 
induces the transformations
\begin{eqnarray}
\mathcal{C} \leftrightarrow \mathcal{C}' + \mathcal{L}, \\
\mathcal{L} \leftrightarrow \mathcal{L}_1 + \mathcal{L}_2.
\end{eqnarray}
In what follows, we will denote each $N$-dipole state as an element of the permutation 
group $\mathcal{P}_N$. For simplicity we suppress the transverse coordinates 
in the notation, 
and each gluon is denoted by a number indicating the order in which 
it was emitted, the first emitted gluon is denoted 1, the second 2 and so on. 
The initial $q\bar{q}$ pair is simply denoted by 0. A generic $N$-dipole 
state containing 
$N-1$ gluons, with $k_0-1$ gluons in the open chain, and the rest in $m$ 
closed loops each containing $k_i$ gluons, is denoted 
\begin{eqnarray}
\mathscr{A}_{N}=(0\, \alpha_1\dots \alpha_{k_0-1})(\alpha_{k_0}\dots \alpha_{k_0+k_1-1})\dots 
(\alpha_{\sum_{i=0}^{m-1}k_i}\dots \alpha_{N-1}) \in \mathcal{P}_N. \label{eq:dipstate} 
\end{eqnarray}  
\FIGURE[t]{
  \begin{minipage}{15cm}
    \begin{center}
      \epsfig{file=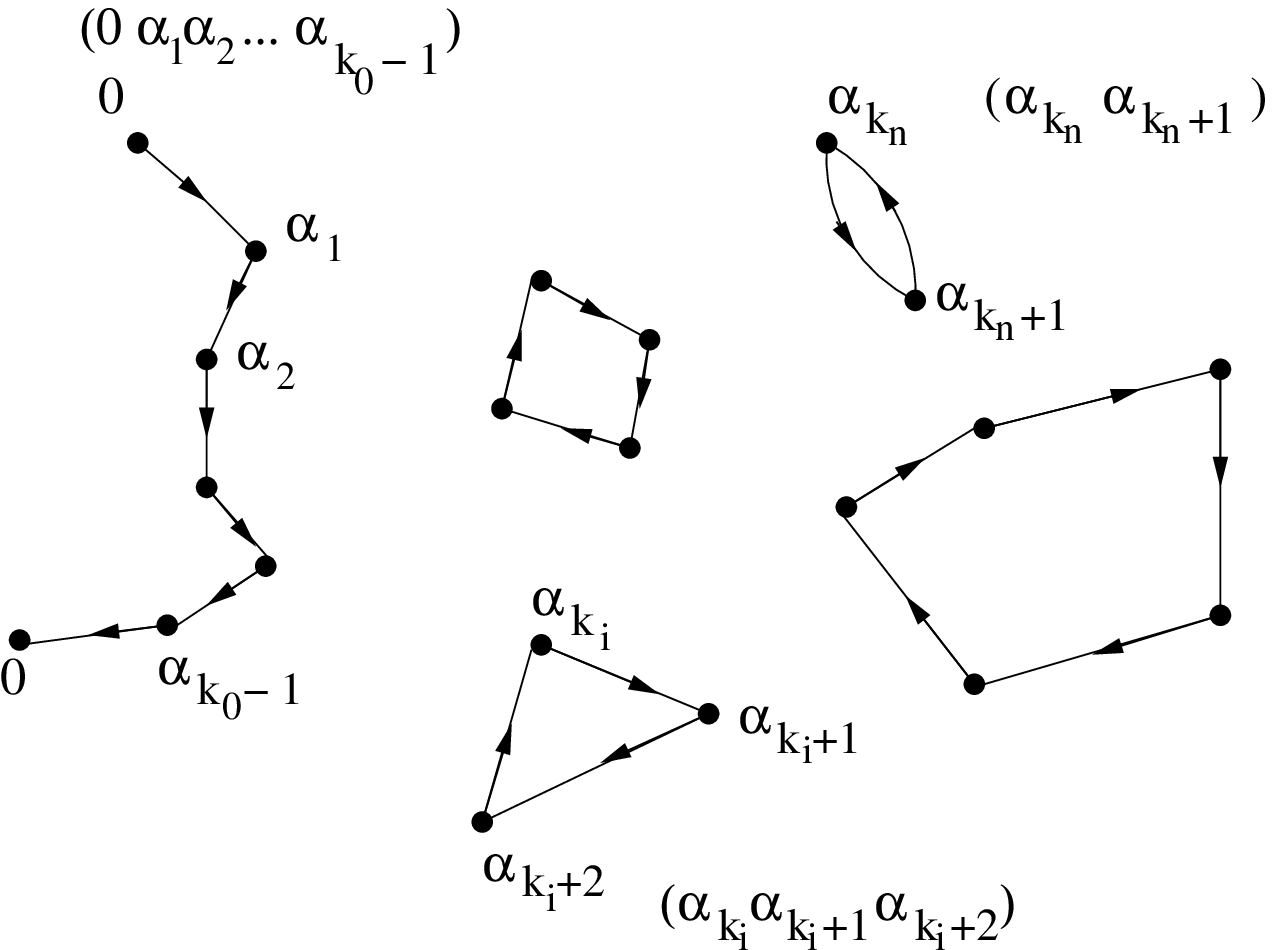,width=8cm}
    \end{center}
  \end{minipage}
  \caption{\label{fig:gendipstate} A generic dipole state formed after 
   a rapidity evolution of $Y$, starting from a $q\bar{q}$ pair. The initial 
   quark and the antiquark are both denoted by 0 while the gluons are 
   denoted by $\alpha_i$ as explained in the text and in eq \eqref{eq:dipstate}. 
   The arrows 
   on the dipoles indicate the colour flow, which goes from colour to anti-colour 
   as before.}
}
Here $\{\alpha_i\}_{i=1}^{N-1}$ is a permutation of $i=1, \dots, N-1$.
The generic dipole state in \eqref{eq:dipstate} is illustrated in figure \ref{fig:gendipstate}. 
Each arrow indicates the colour flow, and in the group theoretical notation 
in \eqref{eq:dipstate}, each gluon $\alpha_i$ points to the gluon to the right 
of it. The open chain is always represented by the cycle 
containing the element 0 (the $q\bar{q}$ pair), and each cycle 
in \eqref{eq:dipstate} corresponds to a colour singlet. Since we cannot 
have colour singlet gluons, the numbers $\alpha_i$ 
cannot appear as 1-cycles. The only 1-cycle allowed is $(0)$, which corresponds to a 
dipole formed by the initial $q\bar{q}$ pair.

Every element in the group $\mathcal{P}_N$ belongs to a certain class, 
which is determined by the cyclic structure of the element. 
The group $\mathcal{P}_4$ has 5 classes: $1111$, $211$, 
$31$, $22$ and $4$. Here each $n$-cycle is represented by the number 
$n$. The state $\mathscr{A}_N$ in \eqref{eq:dipstate} belongs to 
the class $k_0\,k_1\dots k_m$. The identity element is the permutation 
which takes every number onto itself, and has the cyclic structure $11 \dots 1$.

A swing operation can be represented by an element of $\mathcal{P}_N$ which 
consist of one 2-cycle and $(N-2)$ 1-cycles, \emph{i.e.} by an element 
belonging to the class $2 11\dots 1$. 
Thus for example, the swing illustrated in fig \ref{fig:ijswing} 
is represented by $S(\alpha_i,\alpha_j)=(\alpha_i \alpha_j)\prod_{k\neq i,j} (\alpha_k)$, 
and we have
\begin{eqnarray}
S(\alpha_i,\alpha_j)\otimes (\dots \alpha_{i-1} \alpha_i \dots \alpha_{j-1} \alpha_j \dots)&=& 
(\alpha_i \alpha_j)\prod_{k\neq i,j} (\alpha_k)\otimes (\dots \alpha_{i-1} 
\alpha_i \dots \alpha_{j-1} \alpha_j \dots)
\nonumber \\
&=& (\dots \alpha_{i-1} \alpha_j \dots ) (\alpha_{i} \dots \alpha_{j-1}).
\label{eq:ijswing}
\end{eqnarray}
Here $\otimes$ denotes the group multiplication. The action of 
$S(\alpha_i,\alpha_j)$ makes $\alpha_{i-1}$ point at $\alpha_j$, and 
$\alpha_{j-1}$ point at $\alpha_i$, leaving all other $\alpha_k$ unchanged
as shown in the figure. 

\FIGURE[t]{
  \epsfig{file=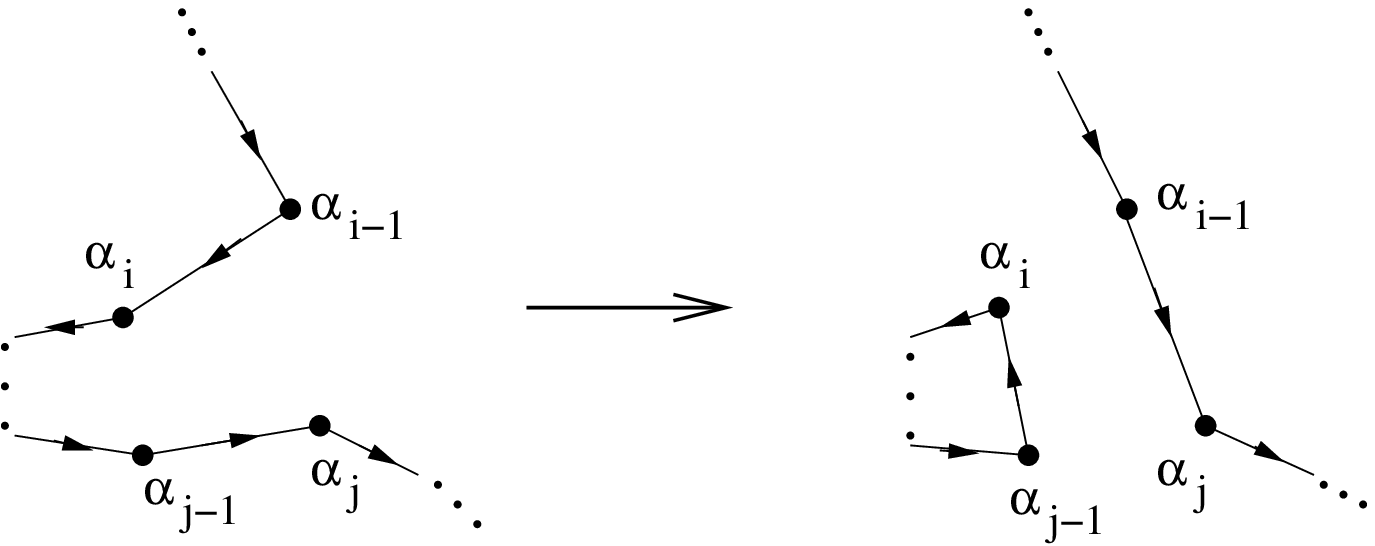,width=9cm}
 \caption{\label{fig:ijswing} Illustration of eq \eqref{eq:ijswing}. }
}

Due to the fact that not every swing leads to a physically acceptable state, 
the number of allowed swings for a state containing $N$ dipoles is not 
simply $\frac{1}{2}N(N-1)$. 
This would e.g. be the number of pairs in a reaction-diffusion type of 
formalism. In most formulations this is not taken into account, 
but we here wish to emphasize the importance of keeping track 
of the correct topology of the evolution. While this is not 
important in the original formulation of the dipole model where the 
dipole state evolves through the $1\to 2$ splitting only, 
it is very necessary for transitions involving more than 
one initial dipole. 
In the appendix we show that the number of physically 
possible states $\mathcal{N}_D$, and the number of possible swings 
$\mathcal{N}_S$ are for $N$ dipoles given by 
\begin{eqnarray}
\mathcal{N}_D(N) &=& (N-1)!\sum_{l=0}^{N-1} \frac{(-1)^l}{l!}(N-l), \label{eq:nodipstate} \\
\mathcal{N}_S(N) &=& \frac{1}{2}(N-1)(N-2)+n_2 \label{eq:noswing},
\end{eqnarray}  
where $n_2$ is the number of closed loops containing 2 dipoles. \\

\emph{Multiple swings in the $N\to N+1$ evolution}
\\

The classes of the group $\mathcal{P}_N$
are connected to each other via the swing as illustrated in fig 
\ref{fig:p4p5classdiag}, where 
each line means that two elements from the respective classes can be transformed 
into one another using one swing. Note that the longest distance is that between 
$4$ and $1111$, which requires 3 swings. In $\mathcal{P}_5$,
we need $4$ swings to go from $11111$ to $5$, as is also shown in fig \ref{fig:p4p5classdiag}. 

Generally, for $\mathcal{P}_N$, any element in 
the class $N$ can be reached from the 
identity element using $N-1$ swings. Explicitely, we 
can write the $N$-cycle $(j_1\dots j_N)$ as 
\begin{eqnarray}
(j_1\dots j_N) = S(j_1, j_N)
%\prod_{k \neq 1,N} (j_k) 
\otimes S(j_1, j_{N-1}) 
%\prod_{k \neq 1,N-1}\!\! (j_k) 
\otimes \dots \otimes S(j_1, j_2)
%\prod_{k \neq 1,2} (j_k).
\end{eqnarray}
This also implies that, 
given any arbitrary element $a \in \mathcal{P}_N$, we can reach any 
other element $b \in \mathcal{P}_N$ using at most $N-1$ swings. This 
is so since we can always find $N-1$ swings such that their 
product equals $ba^{-1}$. 

\FIGURE[t]{
 \begin{minipage}{15cm}
   \begin{center}
    \epsfig{file=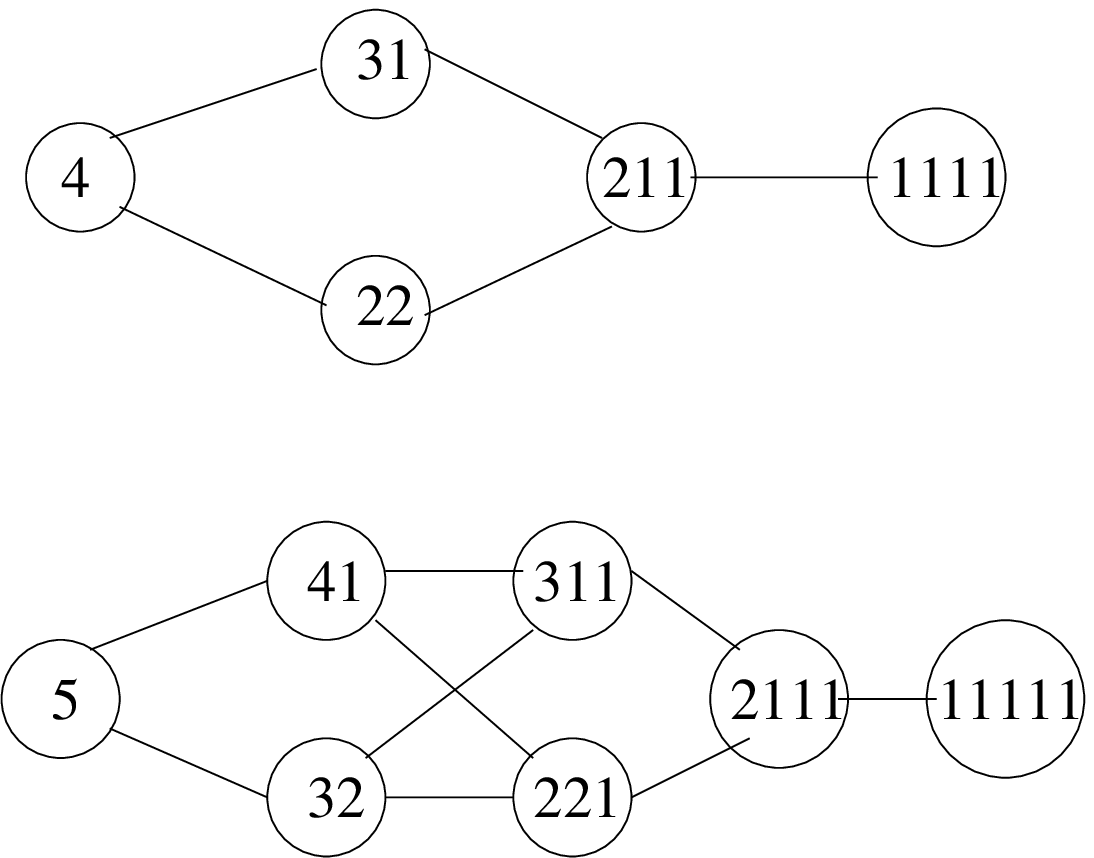,width=0.45\linewidth}
  \end{center}
  \end{minipage}
  \caption{\label{fig:p4p5classdiag} Class diagrams for $\mathcal{P}_4$ and 
    $\mathcal{P}_5$.}
}

\FIGURE[t]{
  \begin{minipage}{15cm}
    \begin{center}
    \epsfig{file=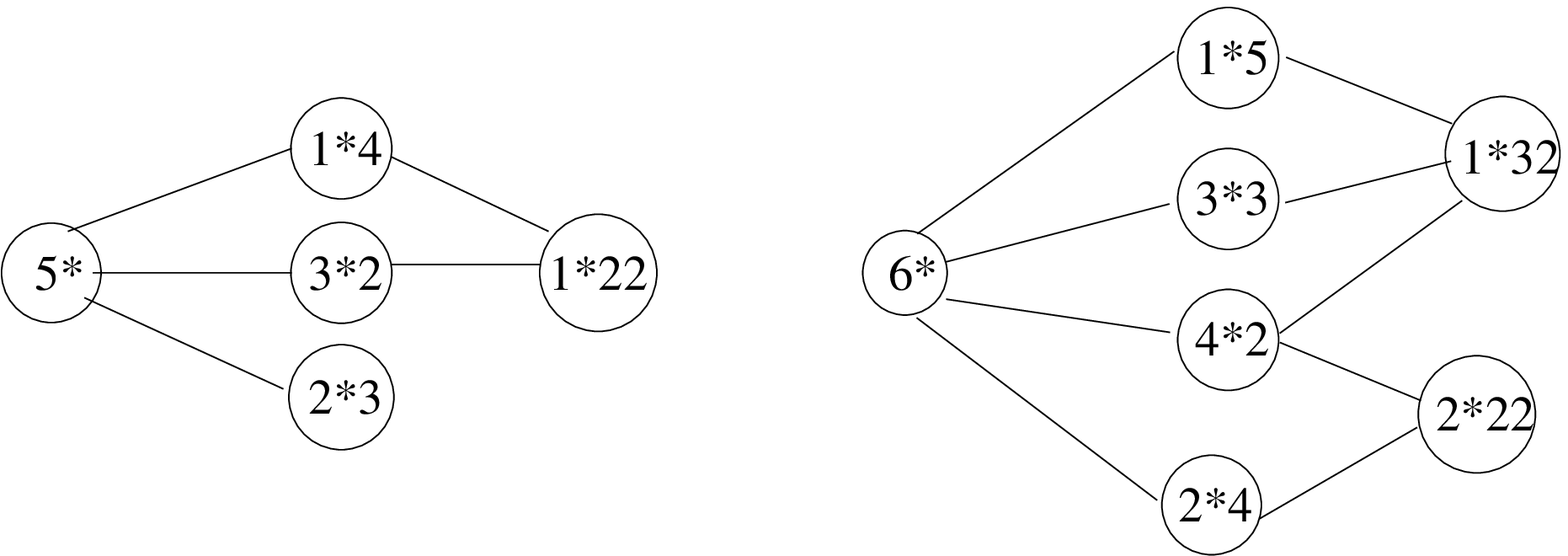,width=0.8\linewidth}
  \end{center}
  \end{minipage}
 \caption{\label{fig:p5p6classdiag} Class diagrams for the 
   subset of physical states of $\mathcal{P}_5$ and 
   $\mathcal{P}_6$. The open chain is marked by $*$.}
}

However, not all classes fall into the subset of physically acceptable states,  
which in particular 
does not contain the identity element. Therefore we cannot a priori 
say whether or not the result above also holds for this subset. In fig 
\ref{fig:p5p6classdiag},
we show the class diagrams of physically acceptable states for 
$\mathcal{P}_5$ and $\mathcal{P}_6$. Here $n^*$ 
denotes the open chain containing $n-1$ gluons. Thus using this notation
we would say that $\mathscr{A}_N$ in \eqref{eq:dipstate} belongs 
to the ``class'' $k_0^* k_1 \dots k_m$.
With a slight abuse of 
nomenclature, we will for simplicity continue to refer to these 
quantities as ``classes'', even though they do not constitute classes 
in the group theoretical sense.

Actually, the dipole splitting can be represented by the same 
class of elements as the dipole swing. Assume we are in the state
$\mathscr{A}_{N}$. We then regard the splitting as a two-step process;
first, we add the $N$th gluon as a 1-cycle into the state $\mathscr{A}_{N}$, 
formally writing $\mathscr{A}_{N}$ as an element of $\mathcal{P}_{N+1}$,
\begin{eqnarray}
\bar{\mathscr{A}}_{N}=(0\, \alpha_1\dots \alpha_{k_0-1})(\alpha_{k_0}\dots \alpha_{k_0+k_1-1})\dots 
(\alpha_{\sum_{i=0}^{l-1}k_i}\dots \alpha_{N-1})(N) \in \mathcal{P}_{N+1}.
\end{eqnarray} 
We put a bar on $\bar{\mathscr{A}}_{N}$ since, written in this way, it is not a 
physically acceptable state. Then, in the second step, we represent 
the emission of $N$ from the 
dipole spanned between $\alpha_i$ and $\alpha_{i+1}$ by operating on 
$\bar{\mathscr{A}}_{N}$
with $S(N,\alpha_{i+1})$ (see eq \eqref{eq:ijswing}), since in that case $(\dots \alpha_i \, 
\alpha_{i+1}\dots )$ is replaced by $(\dots \alpha_i\,N\,\alpha_{i+1}\dots )$.

\FIGURE[t]{
  \begin{minipage}{15cm}
    \begin{center}
      \epsfig{file=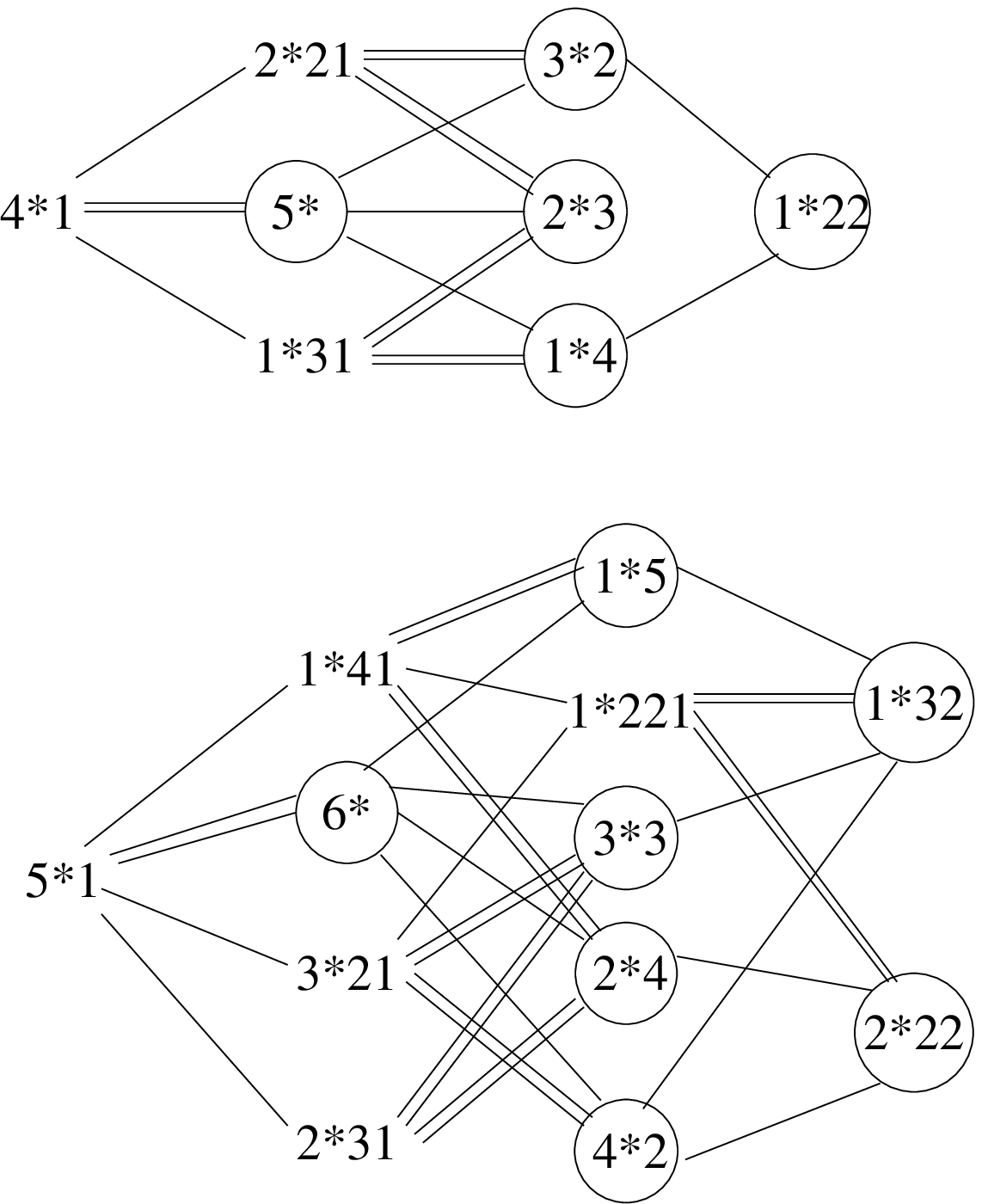,width=0.5\linewidth}
    \end{center}
  \end{minipage}
  \caption{\label{fig:p3p4classdiag} Class diagrams representing 
   the evolutions $N=4 \to N=5$, and $N=5 \to N=6$. Here the double 
   lines represent
   the dipole splitting. Note that for these we can only go in 
   one direction, from an un-circled class to a circled one.}
}

The class diagrams for the generic $N \to N+1$ evolution 
can be drawn in a similar fashion 
as before. In fig \ref{fig:p3p4classdiag}, we show examples 
for $N=4$ and $N=5$. Here only 
the circled classes are physically acceptable, and it is one of these 
that we must end up in, starting from one of the un-circled ones. 
The maximal distance between any two circled classes in an $N\to N+1$ 
evolution is 2 for $N=3$, 3 
for $N=4$, and also 3 for $N=5$. Thus in this case this distance 
is not equal to $N-1$ for a $N$-dipole state. This does, however, 
not automatically imply that we can reach any given state in less 
than $N-1$ swings. 

We here conjecture that 
one can also for the subset of physical states 
go from $\mathscr{A}_N$ to any $\mathscr{A}_{N+1}$ 
by combining a dipole splitting with at most $N-1$ swings. 
In the appendix we show explicitely that this statement 
is true for $N=4$ (the cases $N=1, 2$ are trivial, and $N=3$ can 
be checked very easily). We have also 
checked this result for $N=5$, $N=6$ and $N=7$, but we will for simplicity 
not present the calculations for these cases.  
In a process 
where $k-1$ swings take place, $k$ dipoles in the cascade get replaced
by $k+1$ dipoles, thus giving a $k\to k+1$ transition. 

The
only exception to the statement above is when there are states containing 
an isolated triangle. For example, if we wish to go from the state 
$(0)(1\,2\,3)(4)$ in $1^*31$, to the state $(0\,4)(1\,3\,2)$ in $2^*3$,
we need 5 steps totally, 1 splitting and 4 swings. Thus for these 
states, $N-1$ swings are not enough (one needs $N+1$ swings). 
This is directly related to the fact that the physical 
set of states does not allow the steps: $(1\,2\,3) \to 
(1)(2\,3) \to (1\,3\,2)$. Therefore we need to use 4 swings rather 
than only 2 swings, which implies that we generally need $N+1$ 
swings for these states. 
Note, however, that this problem does not appear for higher 
order cycles. The 4-cycle $(1\,2\,3\,4)$ can for example be 
transformed into $(1\,4\,3\,2)$ easily: $(1\,2\,3\,4) \to (1\,4)(2\,3) 
\to (1\,4\,3\,2)$.  

However,  
we also note that the only difference between the configurations 
$(1\,2\,3)$ and $(1\,3\,2)$ is in the orientation of the dipoles. 
Moreover, the states $\mathscr{A}_N=(\alpha_1\,\alpha_2\,\alpha_3)\mathscr{B}_{N-3}$
and $\mathscr{A}'_N=(\alpha_1\,\alpha_3\,\alpha_2)\mathscr{B}_{N-3}$ have 
exactly the same weights in the cascade evolution. 
They are therefore always produced 
equally, and it is therefore not a problem if we cannot go
between them using $N-1$ swings. 
Finally, we note that the semi-classical approximation 
represented by the cascade evolution cannot take into 
account all quantum-mechanical interference effects.  
The quantum-mechanical states 
corresponding to the configurations $(\alpha_1\,\alpha_2\,\alpha_3)$ and 
$(\alpha_1\,\alpha_3\,\alpha_2)$ have colour factors Tr$(T^aT^bT^c)$
and Tr$(T^aT^cT^b)$ respectively, and for finite $N_c$ these states are not orthogonal. 
Although the interference is suppressed by $1/N_c^2$, 
it is enhanced in for example the decay process $\Upsilon \to 3g$, 
and is in this case quite large \cite{Gustafson:1982ws}, 
which is also confirmed experimentally.

\section{Conclusions}
\label{sec:conclusions}

Mueller's dipole model gives a simple picture of the small-$x$ evolution which is also  
very suitable to use in a MC simulation. While it is known that it 
gives the correct evolution for dilute systems, a fully consistent version for 
dense systems, where saturation
effects during the evolution cannot be neglected, is not known. 
There have been some attempts to interpret these saturation 
effects in terms of dipole mergings but it has not been possible to 
present a consistent probabilistic formulation.  

A consequence of neglecting the saturation effects during the dipole evolution 
is that the model is not frame independent.  
In a previous paper \cite{Avsar:2006jy} (see also \cite{Avsar:2007xg} 
for a more detailed account) we demonstrated that approximate frame 
independence can be achieved by including a so called dipole swing 
in the evolution (the swing was also suggested in \cite{Kozlov:2006cg} 
as a mechanism to generate pomeron loops). Based on this, we constructed 
a phenomenological model 
which, implemented in a MC simulation, gives an almost frame-independent formalism. 

It has been quite difficult to analytically derive the relevant dipole 
interactions which would give rise to saturation effects in the dipole model in 
a way consistent with boost invariance. A very simplified treatment 
of the dipole evolution is offered
by the toy model introduced in \cite{Mueller:1996te, Mueller:1994gb}, 
and later also studied in 
\cite{Kovner:2005aq, Blaizot:2006wp}. In this model it is possible to modify the 
evolution so that the formalism is explicitely frame independent. The 
evolution proceeds here by the addition of a new dipole at each step,
in such a way that the total splitting rate saturates as the dipole occupation 
number gets large. As discussed in \cite{Blaizot:2006wp}, this is actually  
quite similar to the way saturation occurs in the CGC formalism. 
%However, at first 
%sight it seems as if this saturation mechanism is not related to any simple dipole process. 

In this paper we have first shown that it is possible 
to give a probabilistic interpretation to 
the toy model evolutions in terms of positive 
definite $k\to k+1$ transitions. These transitions 
describe the coherent evolution of the dipoles, which is 
an unavoidable consequence of the requirement that 
the transition rates be positive definite. The evolution 
can also be formulated in a more close analogy with a standard 
reaction-diffusion picture, where the $k\to k+1$ transition rates 
only depend on the $k$ dipoles involved in the transition. 
In this case, however, these rates appear with alternating signs which implies 
that a probabilistic treatment is not possible. 

In the real dipole model 
such positive definite vertices can be generated by combining the dipole splitting 
with the dipole swing. In a $k\to k+1$ transition, 
a splitting is combined with $k-1$ simultaneous swings. In the 
approximation where each single dipole only scatters once, we have seen that 
it is enough to combine each splitting with a single 
swing in order to generate the necessary the colour correlations. 
 
When each single dipole is allowed to scatter multiply, one needs
to include more than one simultaneous swing. 
In this case the evolution proceeds by the $k\to k+1$ transitions as in the toy models
mentioned above, and we have further shown that 
for a system of $N$ dipoles,  
one needs at most $N-1$ simultaneous swings in order to 
generate all colour correlations induced by the multiple dipole 
interactions. We therefore obtain a close analogy 
with the toy model evolutions, and the dipole swing furthermore
gives a probabilistic interpretation of the evolution. This 
is easy to show for spatially disconnected dipoles, but it 
is also the case in the more relevant situation when the 
dipoles are connected in chains.  

This statement is strictly speaking not true for states containing a
triangular loop, where only one orientation of this loop 
can be reached using at most $N-1$ swings. This is, however, 
no problem because the two possible orientations always 
appear with the same weight.

\section*{Acknowledgments}

I would like to thank G\"osta Gustafson for valuable discussions and critical reading 
of the manuscript. I am also thankful to Leif L\"onnblad for useful comments, and 
to Bo S\"oderberg for useful discussions on mathematical issues.    

\appendix

\section{The Number of Dipole States}

In this section we will demonstrate that the number of possible states for 
a system containing $N-1$ gluons, together with the initial $q\bar{q}$ pair, 
is given by formula \eqref{eq:nodipstate}. 

We start by considering $n$ gluons in a closed topology, \emph{i.e.} a state 
containing one or more closed dipole loops. If all $n$ gluons are in the same 
loop we obviously have $(n-1)!$ possible states. Next we might have $n$ gluons 
in two loops. The number of such states is given by the number of elements in $\mathcal{P}_n$ 
which consists of one $k$- and one $(n-k)$-cycle. There are $\frac{1}{2}\binom{n}{k}(k-1)!(n-k-1)!$
such elements. The symmetry factor $1/2$ comes from the fact that we can write the $k$-cycle 
either to the left or the right of the $(n-k)$-cycle. For a closed topology consisting of 
$m$ loops, each containing $k_i$ dipoles, the number of possible states is given by 
\begin{eqnarray}
\frac{1}{m!}\prod_{j=1}^{m-1}\binom{\sum_{i=j}^{m}k_i}{k_j}\prod_{i=1}^{m}(k_i-1)!
= \frac{1}{m!}\frac{n!}{\prod_{i=1}^{m}k_i},
\end{eqnarray}  
where
\begin{eqnarray}
\sum_{i=1}^m k_i = n.
\end{eqnarray}
For each fixed closed topology with $n$ gluons we also have $(N-1-n)!$ possible states in the 
open chain. To write down the total number of states it is convenient to introduce a generating 
function $G(z)$ whose series expansion gives the dipole state multiplicity. 
We have 
\begin{eqnarray}
G(z) &=& \binom{N-1}{n}(N-1-n)!n!\sum_{m=0}^\infty \frac{1}{m!}\biggl ( \sum_{k=2}^\infty \frac{z^k}{k}\biggr)^m
\sum_{k_0=0}^\infty z^{k_0}, \nonumber \\
&=& (N-1)!\frac{e^{-z}}{(1-z)^2}
\end{eqnarray}  
where $k_0$ is the number of gluons in the open chain. We also demand that each closed loop 
contain at least 2 dipoles as we do not allow colour singlet gluons. The constraint 
$\sum_{i=0}^m k_i=N-1$ is automatically ensured since we are looking for the $(N-1)$th coefficient 
in the expansion of $G$. The expansion of $G$ gives 
\begin{eqnarray}
G(z) = (N-1)! \sum_{M=0}^\infty\sum_{l=0}^{M} \frac{(-1)^l}{l!}(M-l+1)z^M. 
\end{eqnarray} 
We then immediately see that the $M=N-1$ coefficient is equal to eq \eqref{eq:nodipstate}. Notice also that 
for large $N$ the number of states approaches $\frac{N!}{e}$. This can be 
compared to the number of possible states  
for $N$ dipoles formed by $N$ spatially independent charge--anti-charge pairs, which
is $N!$, and to the number of states in a system consisting of a single  
open dipole chain, as in the original formulation of the dipole model, which is $(N-1)!$.  

\section{The Number of Possible Swings}

In this section we demonstrate that the number of possible swings for a system containing
$N-1$ gluons, together with the initial $q\bar{q}$ pair, is given by eq \eqref{eq:noswing}. 

Assume again that we have $m$ closed loops each containing $k_i$ dipoles 
($i=1, \dots, m$) with $k_i \geqslant 2$. 
The open chain contains $k_0$ gluons, and thus $\sum_{i=0}^m k_i = N-1$. 
Within each closed loop we then have 
\begin{eqnarray}
\frac{1}{2}\sum_{i=1}^m k_i(k_i-3)\theta(k_i \geqslant 3)
\end{eqnarray}
swings. The theta function takes into account the fact that we cannot have any swings in a loop 
containing only two or three dipoles. The number of swings between the closed loops is given by
\begin{eqnarray}
\frac{1}{2}\sum_{i\neq j}^m k_ik_j
\end{eqnarray}
since there are no restrictions in this case. The number of swings between the open chain and the 
closed loops is given by 
\begin{eqnarray}
\sum_{i=1}^m (k_0+1)k_i,
\end{eqnarray}
and finally, the number of swings within the open chain is given by
\begin{eqnarray}
\frac{k_0(k_0-1)}{2}.
\end{eqnarray}

The total number of swings is then given by 
\begin{eqnarray}
&&\frac{1}{2} \biggl \{ \sum_{i=1}^m\{ k_i(k_i-3) - k_i(k_i-3)\delta_{k_i\,2}\}
+ \sum_{i\neq j}^mk_ik_j +  
2(k_0+1)(N-1-k_0) + k_0(k_0-1) \biggr \} \nonumber \\
&=& \frac{1}{2} \biggl \{ (N-1-k_0)^2 - 3(N-1-k_0) + 2n_2 + 2(k_0+1)(N-1-k_0)+k_0(k_0-1) \biggr \} \nonumber \\
&=& \frac{1}{2} (N-1)(N-2)+n_2
\end{eqnarray}
where $n_2$ is the number of closed loops containing 2 dipoles. 

\section{More Details on the $N \to N+1$ Evolution}

In this last appendix, we will explicitely prove that $N-1$ swings are enough 
to reach any arbitrary state for $N=4$. We have also 
explicitely checked the cases $N=5$, but we will for simplicity 
not present these calculations.  We will very briefly  
try to sketch the case when $N=7$. The cases $N=1,2$ are trivial, 
and we also omit the case $N=3$ which is very easy to work out. 

\subsection{$N=4$}

For this case, the class diagram is shown in fig \ref{fig:p3p4classdiag}. Assume first that 
we are in the class $4^*1$. An arbitrary element in this class is given by $(0p_1\,p_2\,p_3)(4)$,
for some permutation $\{p(i)\}$. We must reach any arbitrary 
element using at most 3 swings, and thus using at most 4 steps, counting the 
splitting as one step. 

$\pmb{4^*1\to 1^*22}$: We see from fig \ref{fig:p3p4classdiag} that we have to reach any 
element in $1^*22$ 
using at most 3 steps (or else we would need at least 5 steps). An arbitrary element in  $1^*22$
can be written $(0)(p'_1\,p'_2)(p'_3\,p'_4)$ where $\{p'(i)\}$ is some other permutation. 
Without any loss of generality we might as well assume $p'_4=4$. Then we can always 
start by putting $4$ next to $p'_3$ in the step $4^*1 \to 5^*$. In the next step we can then 
always isolate $(p'_34)=(p'_3p'_4)$ in a 2-cycle. We then have an element 
$(0\pi(1)\pi(2))(p'_3\,p'_4)$ where $(\pi(1),\pi(2))=(p'_1,p'_2)$ or $(p'_2, p'_1)$. Finally 
we can separate$(0\pi(1)\pi(2)) \to (0)(\pi(1)\pi(2))$ to reach $(0)(p'_1\,p'_2)(p'_3\,p'_4)$.
To summarize, we can go through the following steps
\begin{equation}
 4^*1 \to 5^*\to 3^*2 \to 1^*22,
\end{equation} 
and reach any element in $1^*22$ using at most 3 steps. 

$\pmb{4^*1 \to 3^*2}$: Here we can go in either 2 or 4 steps. We then 
want to go to an element $(0p'_1\,p'_2)(p'_3\,p'_4)$. Again we start by putting $4$  
to the right of the number which appears to the left of it in $(0p'_1\,p'_2)(p'_3\,p'_4)$. 
If this number is $0$ (\emph{i.e.} if $p'_1=4$), we can go from $5^*$ to $2^*3$, 
putting 0 and 4 in $2^*$. Then we just go back to $5^*$ to obtain $(0p'_1\,p'_2\,p'_3\,p'_4)$
or $(0p'_1\,p'_2\,p'_4\,p'_3)$. In either we case we can split this chain 
into  $(0p'_1\,p'_2)(p'_3\,p'_4)$. Thus we go through
\begin{equation}
 4^*1 \to 5^*\to 2^*3 \to 5^* \to 3^*2.
\end{equation}
If $p'_1 \neq 4$, we go from $5^*$ to $3^*2$, isolating 4 and its partner in $2$. 
If the three elements in $3^*$ are not in the right order, we can 
split $3^*$ into $1^*2$ and then go back to $3^*$.  
Thus we can go through the steps
\begin{equation}
 4^*1 \to 5^*\to 3^*2 \to 1^*23 \to 3^*2.
\end{equation}

$\pmb{4^*1 \to 2^*3}$: We here want to go to the element 
$(0p'_1)(p'_2\,p'_3\,p'_4)$. Again we start by putting $4$ together with its final partner
($4$ is always put to the right of its partner). If its partner is 0, we can first
go from $5^*$ to $3^*2$ and then go back to $5^*$ so that the elements $(p'_2\,p'_3\,p'_4)$
have the correct permutation. Then we can in one step go the final configuration, 
and thus complete the process 
\begin{equation}
 4^*1 \to 5^*\to 3^*2 \to 5^* \to 2^*3.
\end{equation}
If 4 is in the 3-cycle in the final element, we can go from $5^*$ to $3^*2$ and then 
to $5^*$ again to put all the elements together in the correct positions. This 
is possible since all three elements in the final 3-cycle must have 
the correct permutation. Then we can finish by going from $5^*$ to $2^*3$. 
Thus we can choose the path 
\begin{equation}
 4^*1 \to 5^*\to 3^*2 \to 5^* \to 2^*3.
\end{equation}
Here the 2-cycle in $3^*2$ contains 4 and its partner.

$\pmb{4^*1 \to 1^*4}$: Here we want to reach an element $(0)(p'_1\,p'_2\,p'_3\,p'_4)$. 
Since we always put 4 (without loss of generality we can assume $p'_4=4$) 
next to its partner in the first step, all we need to do is to isolate them ($p'_3$ and $p'_4$) 
in a 2-cycle by going from $5^*$ to $3^*2$. Then we go to $1^*22$, after which we can simply 
join $22$ to $4$, to obtain any desired state. We thus have the path
\begin{equation}
 4^*1 \to 5^*\to 3^*2 \to 1^*22 \to 1^*4.
\end{equation}

$\pmb{4^*1 \to 5^*}$: We can here use at most 3 steps. The final element we want 
to reach has the form $(0p'_1\,p'_2\,p'_3\,p'_4)$. There are two cases, either 4 and 
its partner are linked to 0, or they are not. If they are, we can split $5^*$ into 
$3^*2$ where $3^*$ contains 0, 4 and its partner. Then the other two elements 
can always be put back in $5^*$ in the right position, so that we reach any $5^*$ 
element by 
\begin{equation}
 4^*1 \to 5^*\to 3^*2 \to 5^*.
\end{equation}
In the second case, 4 and its partner are not linked to 0 
(they are $p'_2$ and $p'_3$). Then we can split 
$5^*$ into $2^*3$ where 3 contains 4, its partner and one of the other two 
elements. They will automatically have the correct permutation, and we 
can then get the desired state by joining $3$ and $2^*$ into $5^*$. Then 
we have used the path 
\begin{equation}
 4^*1 \to 5^*\to 2^*3 \to 5^*.
\end{equation}
We have thus seen that we can reach any arbitrary state in $N=5$ from $4^*1$ 
by combining at most 3 swings with a splitting. 

Below we list the cases where we start from $2^*21$. 
In this case we have an initial element $(0p_1)(p_2\,p_3)(4)$. 
By using a splitting first, we can either go to $3^*2$, or to $2^*3$. 

$\pmb{2^*21 \to 5^*}$: First we fix 4 and its partner as usual. If the partner is 
$p_1$, we can separate $(0p_1\,4)$ into $(0)(p_1\,4)$, and then we can 
join $(p_1\,4)$ with $(p_2\,p_3)$ to obtain an element in $1^*4$. Then in 
one step we can go to the desired $5^*$ state. If its partner is 0, and 
the other three elements do not have the correct permutation, 
we can isolate two of them in a 2-cycle (after putting $3^*$ and 2 into $5^*$), 
and then put them back into 
the $5^*$ state in the correct position. Thus we can through the 
two paths
\begin{eqnarray}
&&2^*21 \to 3^*2 \to 1^*22 \to 1^*4 \to 5^*, \\
&&2^*21 \to 3^*2 \to 5^* \to 3^*2 \to 5^*.
\end{eqnarray}
If the final partner of 4 is either $p_2$ or $p_3$, we first 
go to $2^*3$. Then it is easily seen that the two 
paths, 
\begin{eqnarray}
&&2^*21 \to 2^*3 \to 5^* \to 3^*2 \to 5^* \\
&&2^*21 \to 2^*3 \to 5^* \to 2^*3 \to 5^*,
\end{eqnarray}
can take us to any arbitrary element in $5^*$. 

$\pmb{2^*21 \to 1^*4}$: If 4 is next to either $p_2$ or $p_3$, 
we can directly from $2^*3$ go to $5^*$, and then to $1^*4$. If 4 is next to $p_1$, 
we can from $3^*2$ go to $1^*22$ and then in one more step we can reach any 
$1^*4$ state. Thus we can follow the paths 
\begin{eqnarray}
&&2^*21 \to 2^*3 \to 5^* \to 1^*4 \\
&&2^*21 \to 3^*2 \to 1^*22 \to 1^*4,
\end{eqnarray}
to reach any state in $1^*4$ in maximum 3 steps. 

$\pmb{2^*21 \to 1^*22}$: If 4 is next to $p_1$ we can finish 
in 2 steps, $2^*21 \to 3^*2 \to 1^*22$. If 4 is next to either $p_2$ or $p_3$, we can 
first join $2^*3$ into $5^*$, and then isolate 
4 and its partner in the 2-cycle in $3^*2$. Then we need 
only one more step. We thus have the steps
\begin{eqnarray}
2^*21 \to 2^*3 \to 5^* \to 3^*2 \to 1^*22.
\end{eqnarray}

$\pmb{2^*21 \to 3^*2}$: If $p_1$ is partner to 4, we need at most 
go to $1^*22$ from $3^*2$, and then back to $3^*2$ to finish. If 4 
is next to 0, we can first go to $5^*$ from $3^*2$, and then 
split $5^*$ into the desired $3^*2$ state. If 4 is next to $p_2$ or $p_3$, 
we can again go to $5^*$ and then directly to $3^*2$. Thus 
we have the steps
\begin{eqnarray}
&&2^*21\to 3^*2\to 1^*22 \to 3^*2 \\
&&2^*21\to 3^*2\to 5^* \to 3^*2 \\
&&2^*21\to 2^*3\to 5^* \to 3^*2.
\end{eqnarray}

$\pmb{2^*21 \to 2^*3}$: If 4 is in the final 2-cycle (\emph{i.e.} next to 0),
all we need to do is to join $3^*2$ into $5^*$, after which 
we can extract the final 3-cycle in one step. If one the other 
hand 4 is in the final 3-cycle, we are after one step 
either finished, or we can from $2^*3$ go to $5^*$, putting 
0 and its final partner together, after which we can split
$5^*$ split into $2^*3$, obtaining the desired state. Thus 
we can go choose one of the paths,
\begin{eqnarray}
&&2^*21\to 3^*2\to 5^* \to 2^*3 \\
&&2^*21\to 2^*3\to 5^* \to 2^*3.
\end{eqnarray}

Finally, we check the case when we start from $1^*31$. 

$\pmb{1^*31 \to 1^*22}$: Here  
we only need two steps: 
\begin{equation}
1^*31\to 1^*4\to 1^*22
\end{equation}
which can be easily seen.

$\pmb{1^*31\to 1^*4}$:  This case is almost trivial, and we can see that 
we 
need at most three steps: 
\begin{equation}
1^*31\to 1^*4\to 1^*22\to 1^*4. 
\end{equation}

$\pmb{1^*31\to 3^*2}$: If 4 appears in the 2-cycle in $3^*2$, 
all we need is to take the steps 
\begin{equation}
1^*31\to 1^*4\to 1^*22\to 3^*2. 
\end{equation}
In the second step we here isolate 4 and its partner in 
one of the 2-cycles. If 4 instead appears in $3^*$, we can 
go through either $1^*31\to 1^*4\to 5^*\to 3^*2$, or $1^*31\to 2^*3\to 5^*\to 3^*2$, 
depending whether or not 4 appears next to 0 in the final configuration. 

$\pmb{1^*31\to 5^*}$: Here we can again have 4 either 
to the right of 0 or not. If not, we just take the steps $1^*31\to 1^*4 
\to 1^*22 \to 3^*2 \to 5^*$. If it is next to 0, we instead take the 
steps $1^*31 \to 2^*3\to 5^*\to 3^*2\to 5^*$. 

$\pmb{1^*31\to 2^*3}$: If 4 appears in $3$, we can just 
go through $1^*31\to 1^*4\to 5^*\to 2^*3$ and finish. However, if 
4 is next to 0, $N-1$ swings are not enough to go to $(04)(p_1\,p_3\,p_2)$ 
as we have already 
discussed in the main text. We have also already noted that this is not 
a problem for the frame independence. With this remark we finish the case 
$N=4$. 
 
\subsection{$N=7$}

\FIGURE[t]{
  \epsfig{file=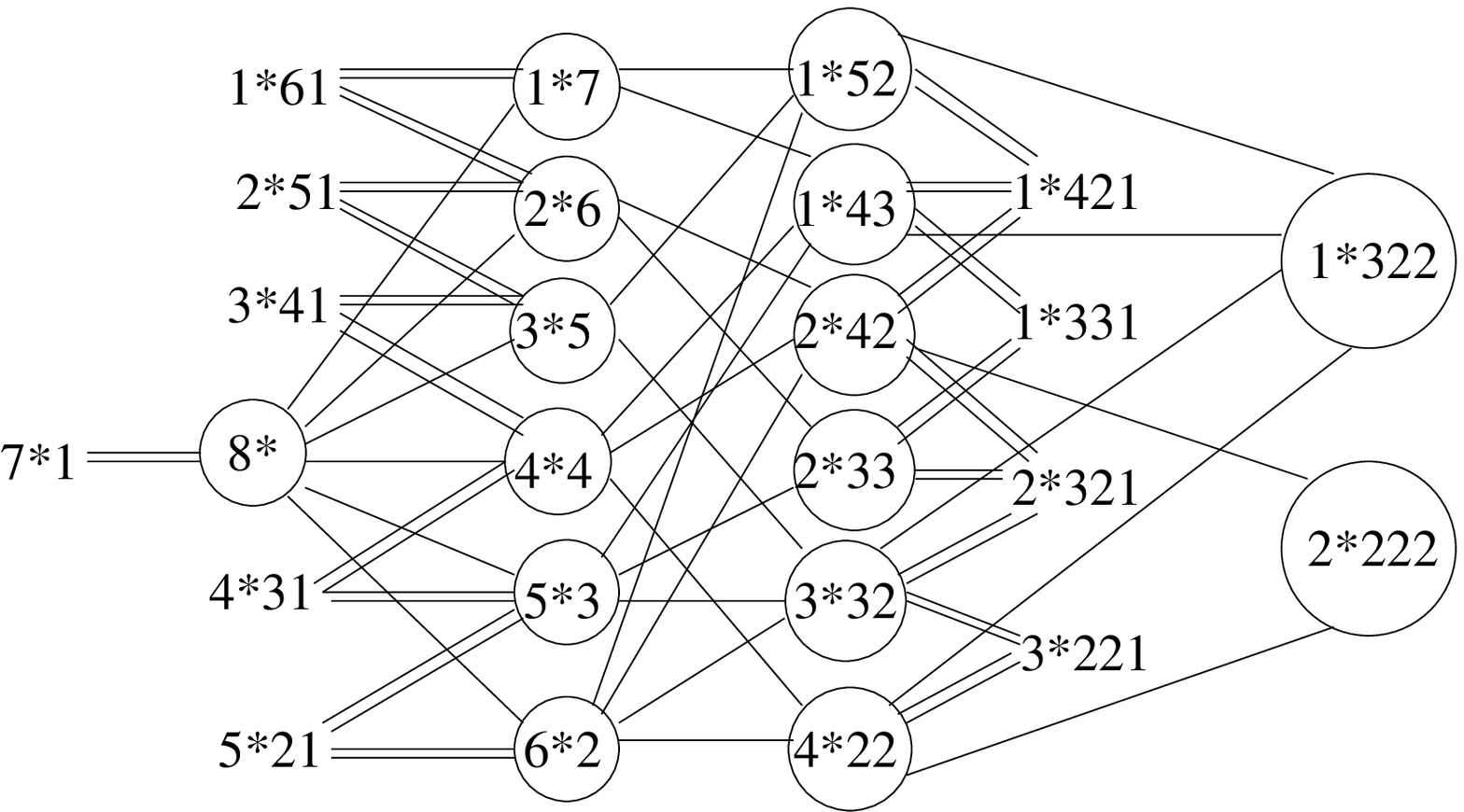,width=10cm}
 \caption{\label{fig:p7p8classdiag} Class diagrams for the 
   evolution $N=7 \to N=8$. Here for simplicity we do not draw
   lines between the un-circled classes.
 }
}

The class diagram for the evolution $N:7\to 8$ is shown in fig 
\ref{fig:p7p8classdiag}. Here it is obviously too tedious to explicitely 
check all possible connections. However, one can now use the results 
from the previous cases to simplify the analysis. For example, 
let us consider the case where we want to go from $7^*1$ to $1^*322$. 
If 7 appears in the final 2-cycle, we can pull it out together 
with its partner after 2 steps, and the question is then whether 
we can go from $6^*$ to $1^*32$ in 5 steps, and we know from 
the case $N=5$ that 
this is indeed possible. If 7 appears in the final 3-cycle, we can after 
4 steps isolate it with its partners, and then we need to go from 
$5^*$ to $1^*22$ in 3 steps, which we also know is possible. Another 
example is if we want to go to a final configuration in $2^*42$, where 7 
appears in 2. After 2 steps, 7 and its partner can again be isolated, 
and we then need to go from $6^*$ to $2^*4$ in 5 steps. Again we know 
that this is indeed possible. We can similarly work out the rest of the cases.
It is also interesting to note that we can here go from $1^*331$ to $2^*33$ 
in 7 steps, even if the initial and final states differ by the 
orientations of the two triangular loops. 

\bibliographystyle{utcaps}
\bibliography{/home/shakespeare/people/leif/personal/lib/tex/bib/references,refs}

\end{document}